\documentclass[]{aa}  
\usepackage{graphicx}
\usepackage{txfonts}
\usepackage{natbib}
\def\deg{$^\circ$}
\def\arcsec{"}

\def\aap{A\&A}

\def\apjl{ApJL}

\def\kms{km~s$^{-1}$}
\def\lsun{L$_{\odot}$}

\def\msun{M$_{\odot}$}
\def\msunyr{\msun\,yr$^{-1}$}
\def\lesssim{\mathrel{\hbox{\rlap{\hbox{\lower4pt\hbox{$\sim$}}}\hbox{$<$}}}}
\def\gtrsim{\mathrel{\hbox{\rlap{\hbox{\lower4pt\hbox{$\sim$}}}\hbox{$>$}}}}

\begin{document}
%


\authorrunning{O. Chesneau et al.}
\titlerunning{The H$\alpha$ line-formation region of Deneb and Rigel }

\title{Time, spatial, and spectral resolution of the H$\alpha$ line-formation region of Deneb and Rigel 
with the VEGA/CHARA interferometer
  \thanks{
    Based on observations made with the CHARA array
  }
}

\author{O. Chesneau \inst{1}, 
  L. Dessart\inst{2}
	D. Mourard \inst{1},  
	Ph. B\'{e}rio\inst{1},
	Ch. Buil \inst{3},
	D. Bonneau \inst{1}, \\
	M. Borges Fernandes\inst{1, 9},
	J.M. Clausse\inst{1},
	O. Delaa\inst{1},
	A. Marcotto\inst{1},
	A. Meilland\inst{4},
	F. Millour\inst{4},
	N. Nardetto\inst{1},\\
	K. Perraut \inst{5},
	A. Roussel\inst{1},
	A. Spang\inst{1},
	P. Stee \inst{1},  
	I. Tallon-Bosc\inst{6}, 
	H. McAlister\inst{7,8}, T. ten~Brummelaar\inst{8}, J. Sturmann\inst{8}, L. Sturmann\inst{8}, N. Turner\inst{8}, C. Farrington\inst{8} and P.J. Goldfinger\inst{8}
	}
	   \offprints{Olivier.Chesneau@oca.eu}

\institute{
UMR 6525 H. Fizeau, Univ. Nice Sophia Antipolis, CNRS, Observatoire de  la C\^{o}te d'Azur, Av. Copernic, F-06130 Grasse, France
       \and
Laboratoire d'Astrophysique de Marseille, Universit\'e de Provence,
CNRS, 38 rue Fr\'ed\'eric Joliot-Curie, F-13388 Marseille Cedex 13,  
France 
 \and
 Castanet Tolosan Observatory, 6 place Clemence Isaure, 31320 Castanet Tolosan, France
 \and
 Max-Planck Institut f\"ur Radioastronomie, Auf dem H¨ugel 69, 53121, Bonn, Germany
 \and
 Laboratoire d'Astrophysique de Grenoble (LAOG), Universit\'e Joseph-Fourier, UMR 5571 CNRS, BP 53, 38041 Grenoble Cedex 09, France
 \and
Univ. Lyon 1, Observatoire de Lyon, 9 avenue Charles Andr\'e, Saint-Genis
Laval, F-69230, France
\and
Georgia State University, P.O. Box 3969, Atlanta GA 30302-3969, USA
 \and
CHARA Array, Mount Wilson Observatory, 91023 Mount Wilson CA, USA
\and
Observat\'orio Nacional, Rua General Jos\'e Cristino, 77, 20921-400, S\~ao Cristov\~ao, Rio de Janeiro, Brazil 
}
\date{Received, accepted.}

\abstract
{BA-type supergiants are amongst the most optically-bright stars. They are observable in extragalactic environments, hence potential accurate distance indicators. 
}
{An extensive record of emission activity in the H$\alpha$ line of the BA supergiants $\beta$\,Orionis (Rigel, B8Ia) and $\alpha$ Cygni (Deneb, A2Ia) 
is indicative of presence of localized time-dependent mass ejections. However, little is known about the  
spatial distribution of these apparent structures. Here, we employ optical interferometry to study the H$\alpha$ line-formation region in these stellar 
environments.
}
{High spatial- ($\sim$0.001$\arcsec$) and spectral- (R=30 000) resolution observations of H$\alpha$ were obtained 
with the visible recombiner VEGA installed on the CHARA interferometer, using     
the S1S2 array-baseline (34m). Six independent observations were done on Deneb over the years 2008 and 2009, and two on Rigel in 2009. 
We analyze this dataset with the 1D non-LTE radiative-transfer code {\sc CMFGEN}, and assess the impact of the wind on the visible and near-IR interferometric signatures,
using both Balmer-line and continuum photons.
}
{We observe a visibility decrease in H$\alpha$ for both Rigel and Deneb, suggesting that the line-formation region is extended ($\sim$1.5-1.75\,$R_{\star}$). 
We observe a significant visibility decrease for Deneb in the Si{\sc ii}\,6371\AA\ line. We witness time variations in the differential phase for Deneb, implying
an inhomogeneous and unsteady circumstellar environment, while no such variability is seen in differential visibilities. 
Radiative-transfer modeling of Deneb, with allowance for stellar-wind mass loss, accounts fairly well for the observed decrease in the H$\alpha$ visibility. 
Based on the observed differential visibilities, we estimate that the mass-loss rate of Deneb has changed by less than 5\%. 
}
{}

\maketitle
%

\section{Introduction}
Supergiants of spectral types B and A (BA-type supergiants) are evolved massive stars of typical initial mass of 25-40\,\msun\ and high luminosity ($\gtrsim$10$^5$\,\lsun). 
Their luminosity and temperature place them among the visually brightest massive stars. Therefore, they are particularly interesting 
for extragalactic astronomy \citep{2008A&ARv..16..209P}. Moreover, they represent attractive distance indicators by means of the
identified wind-momentum-luminosity relationship (WMLR) \citep{2008ApJ...681..269K, 2003MNRAS.345.1223E, 1999A&A...350..970K}. 
As a consequence, nearby BA supergiants have been analyzed with sophisticated radiative-transfer tools. 
Two subjects of intense scrutiny, both observationally and theoretically, have been Deneb ($\alpha$\,Cygni, HD~197345, A2\,Ia) and Rigel ($\beta$\,Orionis, 
HD~34085, B8Ia)
\citep[ for Deneb]{2008A&A...479..849S,  2006ASPC..348..124A, 2002ApJ...570..344A}.

Intensive spectroscopic monitoring of the activity of wind-forming lines, such as H$\alpha$, suggests that the stellar-wind variability 
of luminous hot stars is associated with localized and co-rotating surface structures. The pioneering work 
of \citet{1976ApJ...206..499L} based on a period analysis of radial-velocity curves of Deneb obtained in 1931/32 by \citet{1935LicOB..17...99P} revealed
the simultaneous excitation of multiple non-radial pulsations (NRPs). 
This effort was followed by the extended optical {\sc Heros} campaigns \citep{1998RvMA...11..177K}, monitoring for $\sim$100 consecutive nights 
late B-type/A-type supergiants \citep{1997A&A...320..273K, 1996A&A...314..599K, 1996A&A...305..887K} and early B-type hypergiants \citep{1997A&A...318..819R}. 
Attempts to associate this activity with large-scale surface magnetic fields have so far been unsuccessful 
\citep{2008A&A...483..857S, 2003IAUS..212..255V, 2003IAUS..212..202H, 2003A&A...407..631B}. 
However, the appearance of thin surface convection zones at the supergiant stage may favor the formation and emergence of 
magnetic fields \citep{2009A&A...499..279C}, and may also have a significant impact on potential NRPs.

BA-type supergiants such as Rigel (B8\,Ia) and
  Deneb (A2\,Ia) exhibit observational evidence of the random and pseudo-cyclic activity of the stellar wind.
  A pulsation-driving mechanism has often been proposed, although the associated micro-variability is observed for a wide range 
  of luminosity and $\log T_{\rm eff}$. \citet{1992MNRAS.259...82G} argued for the presence of strange modes. 
\citet{2009A&A...498..273G} tentatively proposed that Deneb's micro-variability is caused by a thin convection zone efficiently trapping 
  the non-radial oscillations, and significant progress has been made in determining the mechanism leading to mass ejection \citep{2009ApJ...701..396C}. 
 The collective effect of multiple NRPs was also proposed as a promising means of explaining the large value of the macro-turbulence parameter; 
  NRPs also represent an attractive mechanism for the formation of Be-star disks \citep{2009A&A...508..409A, 2009A&A...506..143N}.  
  
Based on intensive monitoring of spectroscopic lines, many observational campaigns have attempted to discriminate between these models. Monitoring of Deneb and Rigel 
is still ongoing \citep[ for a non-exhaustive list]{2009NewAR..53..214D, 2008cihw.conf..155M, 2008AstBu..63...23R, 2008A&A...487..211M,
2008A&A...478..823M, 2004MNRAS.351..552M, 1997A&A...318..819R, 1996A&A...305..887K}.
  
The uniform-disk (UD) angular diameters of Deneb and Rigel are 2.4\,mas and 2.7\,mas, 
respectively \citep{2008poii.conf...71A, 2003AJ....126.2502M}, making them good targets for accurate long-baseline optical interferometry. An extensive study of Deneb was performed by \citet{2002ApJ...570..344A}, using different radiative-transfer models (with allowance for the presence of a wind), and constraints from optical interferometry data. Deneb's wind should enhance the limb darkening relative to hydrostatic models that neglect it. 
However, this can only be constrained observationally by using long baselines in the near-IR (longer than 250m) or by observing in the visible, 
thus relaxing the constraints on the baselines by a factor of about 3-4. The H$\alpha$ line is particularly interesting in that context. 
Its large opacity makes it very sensitive to changes at the stellar surface and above, in particular through modulations in stellar-wind properties.
Its extended line-formation region is therefore more easily resolved than the deeper-forming continuum. \citet{2002A&A...395..209D} investigated the possibility of 
monitoring the mass-loss activity in wind-forming lines by means of optical interferometers equipped with spectral devices of sufficient dispersion. 

BA supergiants rotate slowly ($v \sin i$ of about 25-40km\,s$^{-1}$), at least relative to their terminal wind-velocity of $\sim$200-400\,\kms. 
Spectrally resolving the Doppler-broadened Balmer lines thus requires a resolving power as high as $R=$10 000. 
At optical wavelengths, the most conspicuous indicator of mass-loss in early-type supergiants is H$\alpha$, which exhibits the distinctive
P-Cygni profile morphology and has often been a crucial estimator of the mass-loss rate in spite of its documented variability. 
While \citet{2002ApJ...570..344A} was successful in reproducing many observables obtained for Deneb, the modeling of the H$\alpha$ line proved difficult. 
The most significant and persistent discrepancy between the synthetic and observed
profiles has been the depth of the absorption component, which is significantly weaker in the observed spectrum (residual
intensity $\sim$0.6). Furthermore, while the velocity of the absorption component minimum is quite
variable, it is rarely, if ever, shifted blueward by more than $\sim$50\,\kms, which is $\sim$20\% of the terminal velocity.
\citet{2008A&A...479..849S} used a different strategy, using the H$\alpha$ profile as a reference for a detailed modeling of the wind 
characteristics, and reached a more satisfactory solution. Nevertheless, they also reported some remaining discrepancies in matching 
the H$\alpha$ absorption, and proposed additional effects that may need to be considered in the modeling, such as wind structure 
or the influence of a weak magnetic field.

The VEGA recombiner of the CHARA array is a recently commissioned facility that provides spectrally dispersed interferometric observables, with a spectral resolution reaching R =30 000, and a spatial resolution of less than one $mas$. The instrument recombines currently the light from two telescopes, but 3-4 telescope recombination modes are foreseen in a near future. The H$\alpha$ line of bright, slow rotators such as Deneb or Rigel can be isolated from the continuum, and the spatial properties of the line-forming region can thus be studied with unprecedented resolution. Using the smallest baseline of the CHARA array (baseline of 34m), we conducted a pioneering temporal monitoring of Deneb uncovering a high level of activity in the H$\alpha$ line-forming region. Rigel was also observed a few times. 

The paper is structured as follows. In Sect.~\ref{datared} we present the optical interferometry data and their reduction, 
in addition to complementary spectra obtained by amateur astronomers. We then review in Sect.~\ref{sec:diam} 
the previous interferometric measurements aiming to more tightly constrain the diameter of Rigel and Deneb. 
We then apply this extensive record of measurements in the visible and near-IR continuum to interpret semi-quantitatively the spectrally-dispersed measurements in Sect.~\ref{sec:ha}, 
and to elaborate a radiative-transfer model of the stars in Sect.~\ref{sec:model}.

\section{Observations and data processing}
\label{datared}

\begin{table}[htbp]
  \begin{caption}
    {VEGA/CHARA observing logs. 
     \label{tab:log_obs}
    } 
  \end{caption}
  \centering
  \begin{tabular}{lcc cc}
    \hline
    Date & Time & SCI/CAL & \multicolumn{2}{c}{Projected  Baseline} \\
    & & & Length [m] & PA [\deg] \\
    \hline
    2008.07.28 & 06:04 & Deneb & 33.4 & 6.2 \\
    2008.07.28 & 09:01 & Deneb & 32.9 & -18.8\\
    2008.07.28 & 10:51 & Deneb & 26.2 & -46.7\\
        \hline
    2008.07.30 & 05:20 & Deneb & 33.3 & 10.3 \\
    2008.07.30 & 07:42 & Deneb & 32.3 & -9.9\\
    2008.07.30 & 11:32 & Deneb & 30.1 & -37.6\\

    \hline
    2009.07.27$^{\mathrm{b}}$ & 07:51 & CAL1$^{\mathrm{a}}$ & 31.8 & -19\\
    2009.07.27$^{\mathrm{b}}$ & 08:11 & Deneb & 33.3 & -11.1\\
    2009.07.27$^{\mathrm{b}}$ & 08:28 & CAL1$^{\mathrm{a}}$ & 31.4 & -23\\
    \hline
    2009.08.26 & 06:51 & CAL1$^{\mathrm{a}}$ & 31.1 & -26.8\\
    2009.08.26 & 07:06 & Deneb & 33.0 & -18.5\\
    2009.08.26 & 07:25 & CAL1$^{\mathrm{a}}$ & 30.5 & -31.0\\ 

    \hline
    2009.10.01 & 03:56 & Deneb & 33.2 & -14.3\\
    2009.10.01 & 04:34 & CAL1$^{\mathrm{a}}$ & 33.2 & -14.3\\
    2009.10.01 & 05:56 & Deneb & 32.2 & -27.3\\
    2009.10.01 & 11:35 & Rigel & 25.3 & -6.5 \\
    
    \hline
    2009.10.25$^{\mathrm{b}}$ & 02:21 & Deneb & 32.3 & -26.2\\
    
    \hline
    2009.11.17 & 01:35 & Deneb & 32.8 & -20.4\\
    2009.11.17 & 09:48 & Rigel & 27.1  & -20.5 \\

    \hline

 \end{tabular}
 	\begin{list}{}{}
	\item[$^{\mathrm{a}}$] CAL1: HD 184006, A5V, V=3.7, estimated diameter 0.6$\pm$0.05mas from SearchCal@JMMC
	\item[$^{\mathrm{b}}$] Observations performed remotely from France

	\end{list}
 \end{table}

\subsection{VEGA observations}

Deneb and Rigel were observed with the Visible spEctroGraph And Polarimeter \citep[VEGA instrument,][]{2009A&A...508.1073M} 
integrated within the CHARA array at Mount Wilson Observatory \citep[California, USA,][]{2005ApJ...628..453T}. 
Deneb observations were carried out at regular intervals between July 2008 and November 2009. 
Rigel observations were performed in October-November 2009. The VEGA control system can be handled from France remotely. 
A detailed description of this control system is presented in \citet{2008SPIE.7019E..57C}.

The red detector was centered around 656nm, and we made use of the high-resolution mode (R = 30 000). 
For each observation, VEGA recombined the S1 and S2 telescopes forming the S1S2 baseline. 
This is the smallest baseline of the array, for which the extended H$\alpha$ emission is not over-resolved. 
The S1S2 baseline is almost aligned north-south and the projection of the baseline 
onto the sky does not vary much during the night. This is considered an advantage in the context of 
our observations, which are designed to detect the time variability of the H$\alpha$ line-forming region. 
Details of the observations can be found in Table~\ref{tab:log_obs}. On several occasions, more than 
one acquisition of Deneb was performed during the night, providing interesting, albeit limited information 
about the spatial asymmetry of the source. Since our goal was to investigate the spectral 
and spatial properties of the H$\alpha$ line relative to the continuum, emphasis was not placed 
on obtaining very precise calibration of the absolute visibilities. The angular diameters of Rigel and Deneb 
are well known (see Sect.~\ref{sec:diam}), so we relied on the published values to scale the continuum absolute 
visibilities. Moreover, it is difficult to find a suitable bright calibrator when observing at high spectral resolution, 
which is not well-suited to an accurate determination of the absolute visibility which implies a large continuum window. 
Improving upon the current determination of the angular diameter in the visible would require a dedicated observing 
strategy using the medium-resolution mode, larger baselines coupled with a fringe tracking performed in the near-IR. 
This possibility is foreseen in the future. Therefore, calibrator observations were not performed systematically, 
but only to ensure that the instrument was behaving well.\\

\begin{figure}
 \centering
\includegraphics[width=8.6cm]{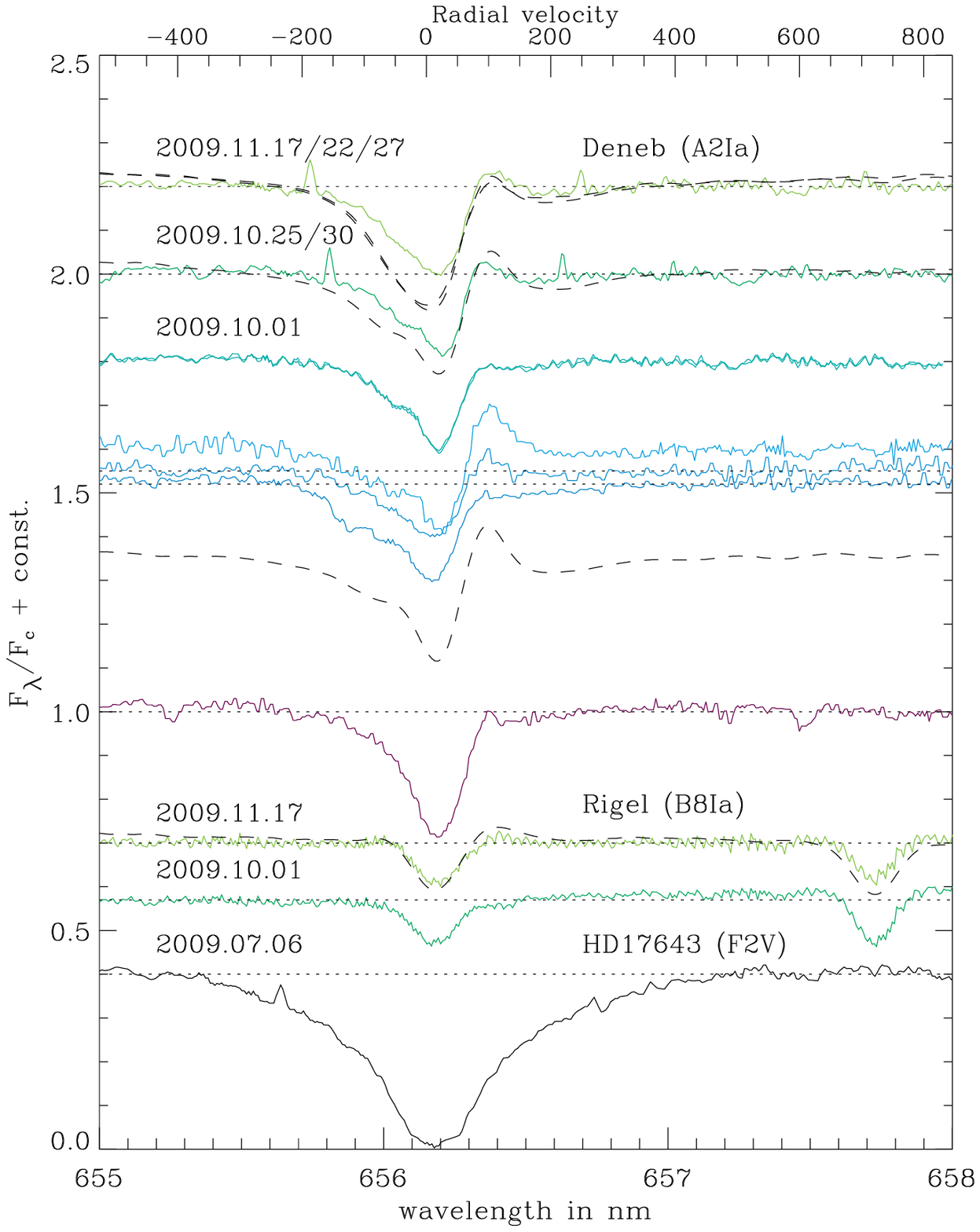}
\includegraphics[width=8.6cm]{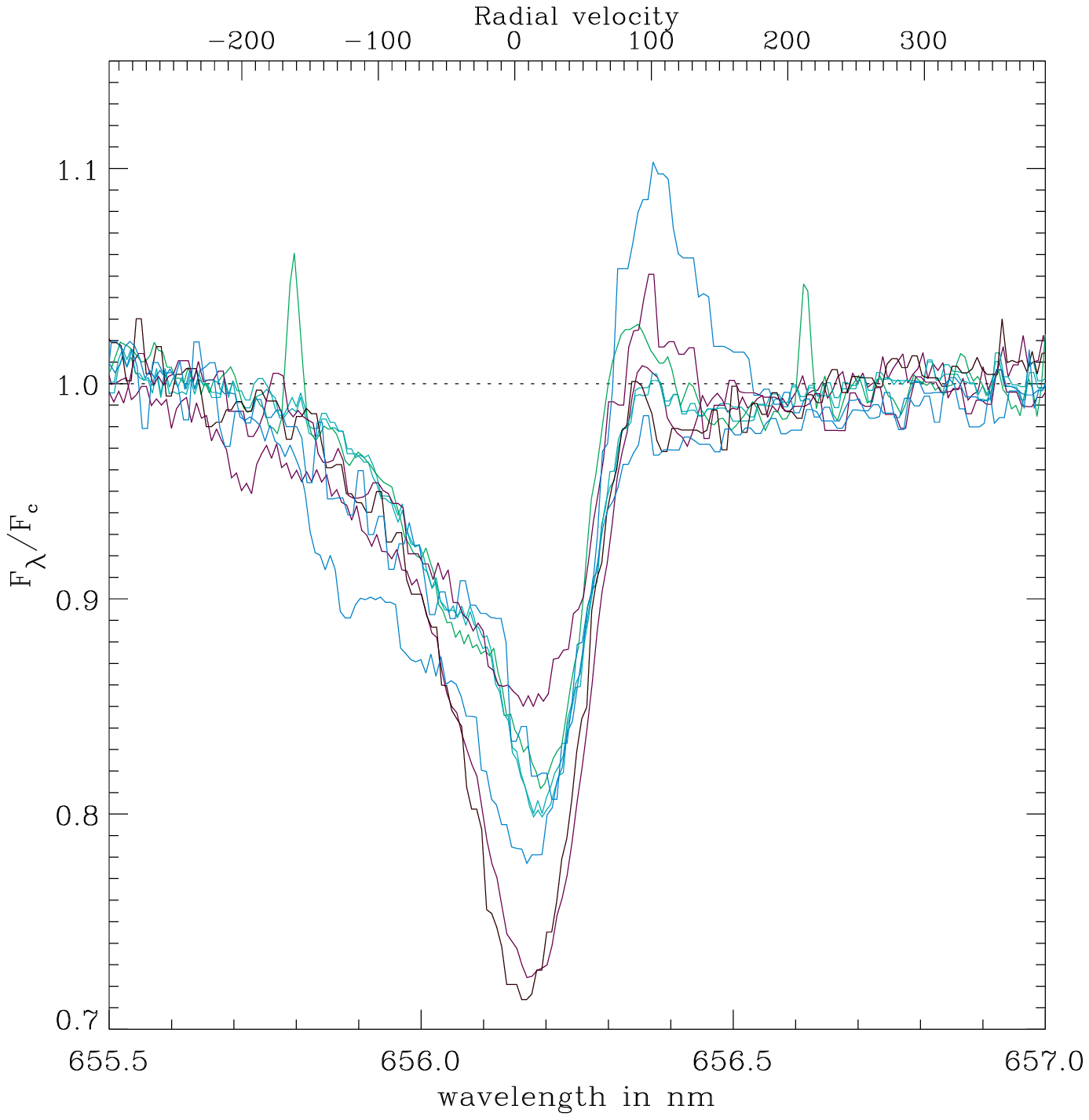}
 \caption[]{
 {\it Top:} Montage of H$\alpha$ observations at various epochs for Deneb and Rigel.
 We also include one observation of the calibrator (HD17643). The dashed lines indicate the spectra recorded by amateur astronomers. 
 {\it Bottom:} Normalized multi-epoch H$\alpha$ observations of Deneb. The color coding is the same as that used 
 in the top panel. \label{fig:spectrum}}
\end{figure}

 The data reduction method is fully described in \citet{2009A&A...508.1073M} and we only summarize it shortly here. 
 The spectra are extracted using a classical scheme, of collapsing the 2D flux in one spectrum, 
 wavelength calibration using a Th-Ar lamp, and normalization of the continuum by polynomial fitting. 
 We note that the photon-counting cameras saturate when the rate of photons is too high locally. Because of the brightness 
 of the sources, neutral density filters of 0.6 to 1 magnitudes had to be applied, depending on the atmospheric conditions and 
 the spread of the speckle images on the slit. Spectra with a signal-to-noise ratio (SNR) of 300-400 were routinely obtained, although clear 
 signs of saturation were found in a few of them. After careful testing, we checked that the saturation had only a very limited effect on the interferometric measurements. 
 When the quality of the spectra was such that telluric lines are observable, these are used to refine the wavelength calibration, 
 although most of the time the VEGA spectral calibration was checked against the reference spectra obtained by amateur astronomers. 
 A series of spectra for Deneb is shown in Fig.~\ref{fig:spectrum}.

The raw squared visibilities (V$^2$) were estimated by computing the ratio of the high frequency energy
to the low frequency energy of the averaged spectral density. The same treatment was applied to the calibrators, whose angular diameter 
was computed using the software SearchCal\footnote{http://www.jmmc.fr/searchcal/} \citep{2006A&A...456..789B}. 
The expected absolute visibilities of VEGA in the continuum from the short baselines do not provide strong 
constraints on the angular diameter and a calibrator was not systematically observed. When such an observation 
is available, we carefully checked that the observed visibilities were compatible with the expected ones. 
A contemporary measurement obtained with the medium resolution mode is presented in \citet{2009A&A...508.1073M}. 

The information in a line was extracted differentially by comparing the properties of the fringe between a reference 
channel centered on the continuum of the source, and a sliding science narrow channel, using the so-called cross-spectrum 
method \citep{1999A&A...345..203B}. The absolute orientation of the differential phase was established by considering the change in the spectral slope of the dispersed fringes \citep{1996ApOpt..35.3002K} when changing the delay line position. Thus a {\it positive} value corresponds to a photocenter displacement along the S1S2 projected baseline {\it in the south direction}.

\begin{figure*}
 \centering
\includegraphics[width=16.cm]{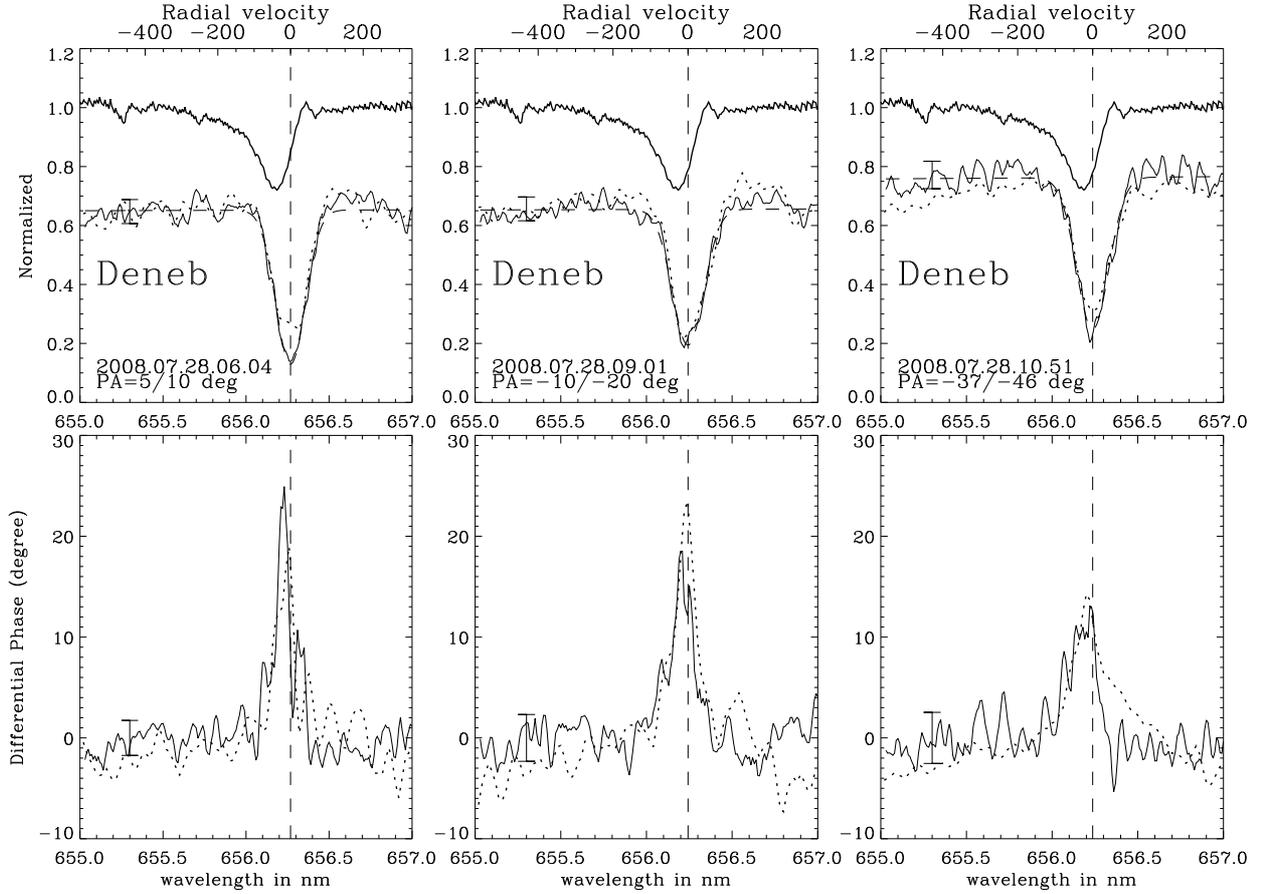}
 \caption[]{ 
{\it Top row:} Normalized flux (upper curve) and dispersed visibility (lower curve; spectral band 0.02 nm) for the H$\alpha$ observations
of Deneb obtained on  2008.07.28 (solid line) and on 2008.07.30 (dotted line). The resolution is $R=30 000$.
{\it Bottom:} Same as top row, but now for the differential phase. A strong signal changing with baseline direction is observed, 
indicating that there was a significant asymmetry in the line-forming region at this time.  \label{fig:vis}
} 
\end{figure*}

\begin{figure*}
 \centering
\includegraphics[width=16.cm]{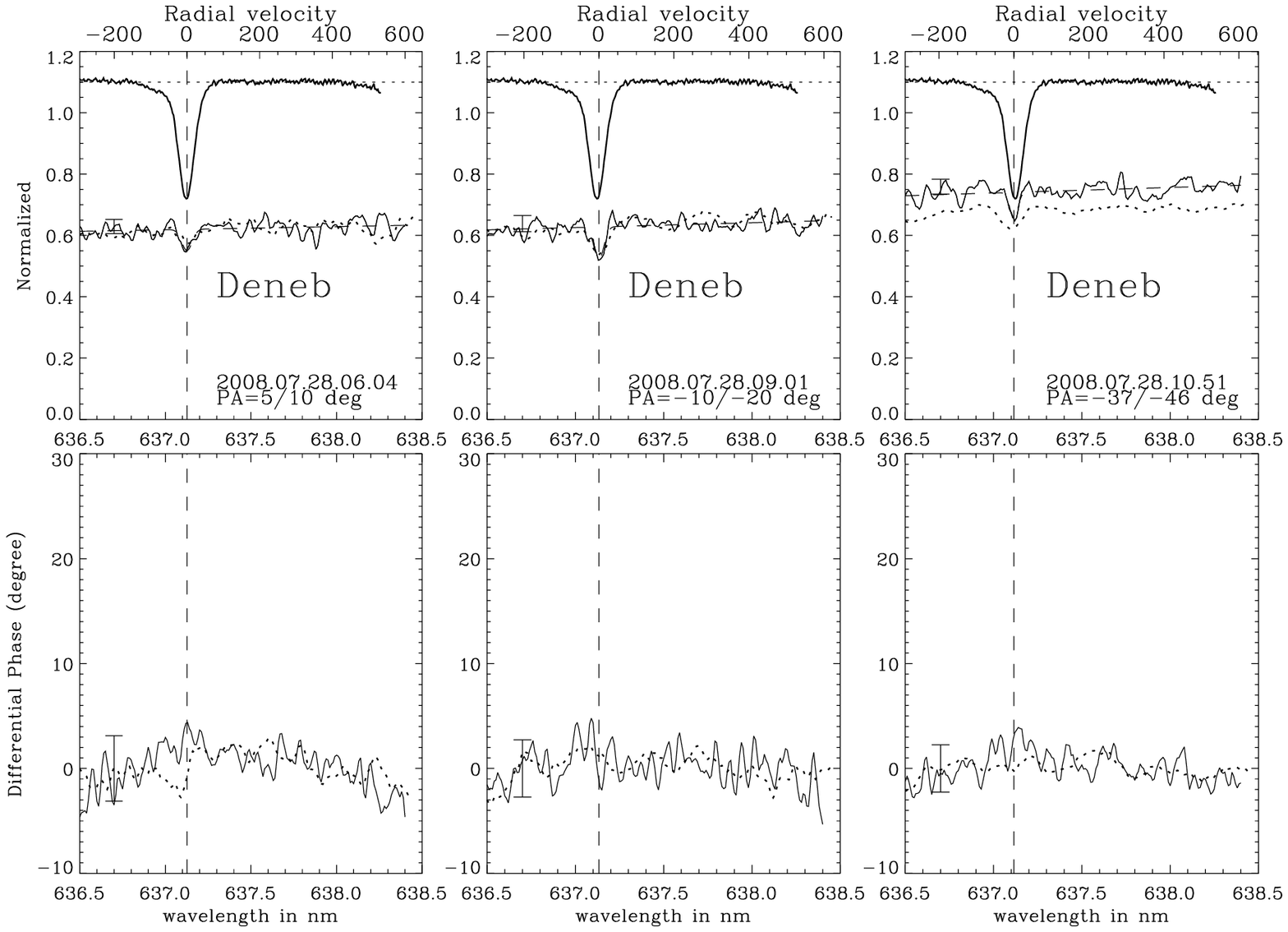}
 \caption[]{\label{fig:visB} 
 Same as Fig.~\ref{fig:vis}, but now showing Deneb observations obtained with the blue camera, whose spectral range covers the Si{\sc ii}\,6371 line. 
 For a clearer comparison, the ordinate ranges are kept identical. 
 The Si{\sc ii}\,6371 line is seen as a strong absorption (the continuum level of the spectrum is offset by 0.1 for clarity),  
 marginally resolved by VEGA/CHARA. However, no phase signal is observed.} 
\end{figure*}

The width of the science channel was 0.02\,nm in good atmospheric conditions, and degraded to 0.08\,nm in poor weather conditions. 
The rms of the spectrally dispersed visibility in the 654-655\,nm continuum ranges from 3-4\% at visibility 1 in the highest quality nights 
(2008/07/28, 2008/07/30, 2009/10/01) to about 7-8\% in medium nights (2009.07.27, 2009.08.26) and more than 10\% 
in poor nights (2009/11/17). The rms of the differential phases follows the same trend ranging from 1-2\deg\ in good 
conditions, to 3-4\deg\ in medium conditions, and 5-6\deg\ in poor conditions. Figures.~\ref{fig:vis}--\ref{fig:visB} 
illustrate the best observations of Deneb secured in 2008. A Gaussian fit was applied to the differential visibilities and 
phase to retrieve accurate information about the spectral FWHM and position of the interferometric signal. The spectral location 
of the differential visibility and differential phase dips are stable at a level of 0.005\,nm ($\sim$3\,\kms). 
Information from the blue camera was also used, as some important lines, e.g. Si{\sc ii}\,6343--6371\AA\ can be investigated 
(see Fig.~\ref{fig:visB} and Sect.~\ref{sec:SiII}). 

\subsection{Spectroscopy from amateur astronomers}

Several H$\alpha$ spectra were obtained during the same period with the 0.28\,m amateur telescope (Celestron 11) 
located in Castanet-Tolosan (France) equipped with the eShel spectrograph and a QSI532 CCD camera (CCD KAF3200ME). 
These spectra were used in this study as an indication of the emission level and variability
of the stars. The typical resolution of these spectra is $\sim$11 000.

The reduction of these data was performed using the standard echelle pipeline (Reshel software V1.11). H2O telluric lines are removed 
by means of division by a synthetic H2O spectrum using Vspec software - the telluric-line list was taken from GEISA database (LMD/CNRS). 
We corrected for the diurnal and annual earth velocity are corrected for (spectral wavelengths are given in an heliocentric 
reference for a standard atmosphere). Systematic differences were found in the VEGA/CHARA and amateur continuum correction 
producing the differences seen in Fig.~\ref{fig:spectrum} between the solid (VEGA) and dashed lines (eShel).

\section{Results}
\subsection{The diameters of Deneb and Rigel}
\label{sec:diam}

For the most accurate estimates to date of the diameters of Rigel and Deneb, we refer to \citet{2008poii.conf...71A, 2006AAS...208.0601A}, 
in which CHARA/FLUOR observations in the K band with baselines reaching 300\,m are described. 
These observations infer a UD angular diameter of 2.76$\pm$0.01\,mas, and 2.363$\pm$0.002\,mas, 
for Rigel and Deneb, respectively. To first order, no significant change in these diameters inferred from the visible 
or the near-IR is expected, the wind being too tenuous to shift the continuum-formation region
(see Sect.~\ref{sec:model}). This value of the Deneb UD angular diameter agrees with the optical measurements 
of \citet{2003AJ....126.2502M} obtained with the MarkIII interferometer using multiple observations with baselines ranging from 3m to 28m with 2.34$\pm$0.05\,mas 
at 0.8\,$\mu$m, 2.26$\pm$0.06\,mas at 0.55\,$\mu$m, and 2.25$\pm$0.05\,mas at 0.41\,$\mu$m. 
The few V$^2$ measurements secured from the present observations with a very limited spatial frequency range agree with the MarkIII measurements. We estimated the uniform disk diameter of Deneb using only the red camera to be 2.31$\pm$0.04mas at 0.65$\mu$m, and using both cameras found the diameter to be 2.34$\pm$0.03mas. Only the best observations could be used, as the visual magnitude of the calibrator (V=3.7) was too faint to often obtain a reliable estimate of the absolute visibility, even using the broadest possible spectral band (i.e. 5nm). 
 
 \begin{table}[htbp]
  \begin{caption}
    {VEGA/CHARA $V^2$ measurements performed on 2009.07.27 and 2009.08.26. 
     \label{tab:log_obs}
    } 
  \end{caption}
  \centering
  \begin{tabular}{lcccc}
    \hline
    Date & Wavelength & Baseline & $V^2$ & $V^2$ error \\
    &[nm] & [m]&  &  \\
    \hline
2009.07.27 & 654 & 32.97 & 0.421 & 0.019  \\
2009.07.27 & 657& 32.97 &  0.444 & 0.024 \\
2009.07.27 & 630 & 32.97 & 0.376 & 0.016 \\
2009.08.26 & 654 & 33.19 & 0.418 & 0.051 \\

    \hline

 \end{tabular}
 \end{table}

\subsection{The H$\alpha$ line-formation region}
\label{sec:ha}

The H$\alpha$ line is one of the most optically thick of all lines seen in the optical and near-IR spectra of hot stars, hence represents an excellent tracer of their winds. This line has mostly been observed by means of 
spectroscopy at  various spectral resolution, and in some dedicated campaigns 
with a very intensive time coverage aimed at recovering the time variability of the line-forming region. 
We refer to \citet{2008AstBu..63...23R} and \citet{2008cihw.conf..155M} for the latest reports on Deneb and Rigel. 
This conspicuous line-profile variability suggests that the wind itself, where the H$\alpha$ line forms, is variable
in its properties. For example, variations may take place in the ionization, the morphology, or the density structure of the wind.
The significant changes in the line-profile shape with time 
indicates that the H$\alpha$ line-formation region is asymmetric, and that this asymmetry changes with time.

The H$\alpha$ line observed in Deneb exhibited evidence of some activity during the 2008-2009 observations, but not of any dramatic change in mass-loss rate. The Rigel spectra were 
also obtained during what appears to have been quiet periods and differ from the ``typical" spectrum shown in 
Fig.~1 of \citet{2008cihw.conf..155M}. As can be seen here in Fig.~\ref{fig:spectrum}, the H$\alpha$ line is as deep 
as the carbon C{\sc ii} lines at 657.8 and 658.3\,nm in Rigel. This is not an unusual state, as H$\alpha$ often appears 
in pure absorption and to be symmetric about the line center (in the rest-frame)
about 20\% of the time (using \citet{2008cihw.conf..155M}  statistics).

Both the Deneb and Rigel dispersed visibilities exhibit a profound dip in the H$\alpha$ line, 
suggesting that the line forms over an extended region above the continuum.
These dips are symmetric about the line center (in the rest-frame) in both objects.
The FWHM of the visibility dip was estimated by performing a Gaussian fit to data of the high quality nights in 2008. 
These measurements yield a FWHM of 0.215$\pm$0.007\,nm, i.e. 98$\pm$3\,\kms\ for Deneb (Fig.~\ref{fig:vis}). 
By comparison, the visibility signal is narrower for Rigel, with a FWHM of 70$\pm$5\,\kms. 
One can estimate the visibility in the line by using a continuum derived from published values 
of the angular diameter. The corresponding UD estimates for the highest extension 
of the H$\alpha$ line-forming regions are 4.1$\pm$0.2\,mas and 4.2$\pm$0.2\,mas for Deneb and Rigel, 
representing 1.75 and 1.5$R_{\star}$, respectively.

\begin{figure}
 \centering
\includegraphics[width=8.5cm]{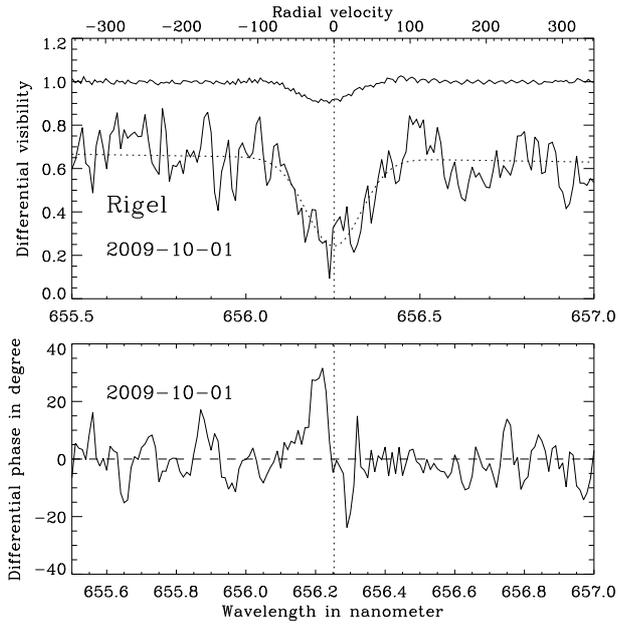}
 \caption[]{{\it Top:} H$\alpha$ observations of Rigel on 2009.10.01 (upper curve), and the corresponding 
 differential visibility, scaled to the expected absolute level of the continuum (lower curve). 
{\it Bottom:} Differential phase curve exhibiting a clear 
S-shape signal, indicative of rotating circumstellar material. A vertical dotted line is shown in both panels to indicate
the location of the H$\alpha$ rest-wavelength determined by a Gaussian fitting of the visibility curve.
\label{fig:rigel}}
\end{figure}

The differential phases are a direct indication of the position of the H$\alpha$ line-forming regions at different radial 
velocities, by comparison to the continuum considered as the reference of phase. 
For the 2008 data, the Gaussian fits provide FWHMs that increase from the baselines oriented near +10\deg\ to 
-40\deg\ from 61$\pm$2\,\kms\ to 84$\pm$4\,\kms. Sub-structures are also detected, with the highest 
peak near the zero velocity, and two satellites at about 40\,\kms. The peak level reached by the 
differential phases follows a trend from a large photo-center shift at P.A. close to 5-10\deg\ that decreases toward $-$30-50\deg.

\begin{figure}
 \centering
\includegraphics[width=8.5cm]{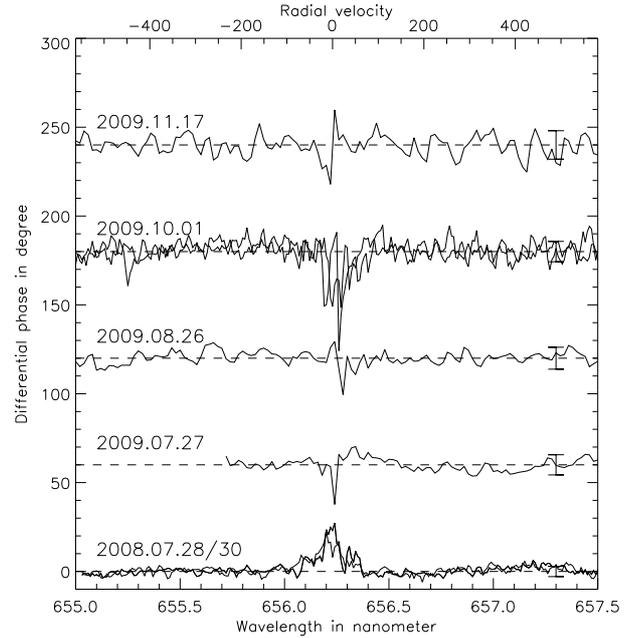}
 \caption[]{Time evolution of the differential phases of Deneb. The strong phase signal observed in 2008 is indicative 
 of a very asymmetric environment, off-centerd from the continuum source. No clear S-shape signal is observed, 
 in agreement with the low $v\sin i$ of Deneb of $\sim$20 km\,s$^{-1}$, unresolved by the instrument. 
\label{fig:phases}}
\end{figure}

The differential phases recorded for Deneb between 2008 and 2009 are shown in Fig.~\ref{fig:phases}. 
Dramatic changes are observed with periods of large phase signal alternating with periods containing no detectable signal. 

The ``calm" periods may potentially provide important information about the rotation of the star.
This characteristic signal can be observed in the rotating circumstellar environment of Be-star disks 
(see Delaa et al. in preparation; \citealt{1999A&A...345..203B,1998A&A...335..261V,1996A&A...311..945S}), 
and also directly on the photosphere of a rotating star \citep{2009A&A...498L..41L}. Approaching and receding regions of
a rotating star experience different Doppler shifts and are thus spectrally separated. 
In the sky, the natural consequence 
is that the line-absorbing regions are seen by the interferometer at one or the other side of the continuum, generating the 
well-known S-shape signal in the differential phases. This signal might have been detected in Rigel data of the 2009-10-01 
(see Fig.~\ref{fig:rigel}). The rms of the phase of these data is 7.7\deg, and the maximum and minimum of the signal are at 26\deg\ and -17\deg, respectively, at the blue and red sides at $\sim$15\,\kms\ from line center. 
The signal is kept at a similar level when extracted with a double binning of 0.04\,nm, and the phase rms is 
decreased to 5.6\deg. This is on the order of the estimated $v\sin i$ of the source of about 36\,\kms. We were 
unable to detect a similar signal for Deneb probably because of the limited spectral resolution, insufficient to resolve 
its low estimated $v\sin i$ of 20\,\kms. As the signal is anti-symmetric, it cancels out if the spectral resolution is too low. 
One must also keep in mind that the quality of the data fluctuated and that the binning could not be kept identical 
at all epochs. It is not impossible that weak signals are blurred (such a signal might be visible the 2009/11/17; 
Fig.~\ref{fig:phases}). 
Given the large extension and the activity observed in the differential phases in H$\alpha$, this line is probably 
not ideally suited to inferring the rotation of the star.

To first order, and for marginally resolved sources, the differential phases can be considered to linearly depend on the 
photo-center position of the emitting source. This is no longer the case when the source is significantly resolved as 
in our case, but we consider this as a rough estimate. Assuming that the line 
emission represent a fraction f$_{ratio}$ of typically 50\% of the light in the H$\alpha$ line core, the following 
relation is used $p=-(\phi/360)(\lambda/B)(1/f_{ratio})$. 
We find that the astrometric shifts induced by the inhomogeneous H$\alpha$ emission reach 0.5\,mas (0.2$R_{\star}$), 
but are lower than 0.3\,mas most of the time. The event observed in 2008 is remarkable as the full line appeared to be 
off-center relative to the continuum. A similar event was observed on 2009/10/01 to have an opposite direction.

\subsection{The Si{\sc ii}\,6347--6371\AA\ line-formation regions}
\label{sec:SiII}

The Si{\sc ii}\,6371 line was observed simultaneously with the H$\alpha$ line in the blue camera (Fig.\ref{fig:visB}) in 2008.
The line profile shows a pure absorption, but is slightly asymmetric with an extended blue wing due to extended absorption in the wind regions.
The VEGA/CHARA observations of Deneb obtained in 2008 indicate that the line-formation region of the Si{\sc ii}\,6371 
line is more extended than the continuum forming region, with differential-visibility dips of 10\%, 19\%, and 15\% for baselines 
in the range [5\deg:10\deg], [-10\deg:-20\deg], and [-35\deg:-50\deg]. 
The rms of the dispersed visibilities is 5\%. The line FWHM estimated from Gaussian fitting is 55$\pm$3\,\kms\ and the FWHM 
of the visibility dip is narrower with a FWHM decreasing from 33$\pm$4\,\kms\ to 24$\pm$4\,\kms\ for P.A. from [5\deg:10\deg] 
to [-35\deg:-50\deg], respectively. One can roughly estimate the extension of the line-forming region using the minimum of the visibility and UD approximation to be 2.6\,mas, 2.7 and 2.75\,mas using the value of 2.34\,mas from NPOI as the diameter at 630\,nm. 
One may speculate whether this trend in the visibilities is permanent or a transient event closely related to the asymmetries 
inferred from the differential phases in the H$\alpha$ line. Because of a spectrograph realignment carried out in 2009, the Si{\sc ii}\,6371 line was not any longer observable 
with the blue camera in 2009, but the Si{\sc ii}\,6347 was well-centered and could be analyzed. No such signal was observed at anytime.

Despite the large depth of the silicon lines, no differential-phase signal was detected above an rms of $\lesssim$2\deg. 
This means that the imprint of the rotation of the star on the Si{\sc ii} line-formation region is not detected in the data, 
probably because of insufficient spectral resolution. Strong Si{\sc ii} lines were also observed in Rigel's spectrum, but neither 
differential visibility nor any phase signal was detected. The SNR of the data is poorer than the Deneb observations, 
with an rms in the dispersed visibilities and phases of 10\%.


   \begin{figure*}
 \centering
\includegraphics[width=8.cm]{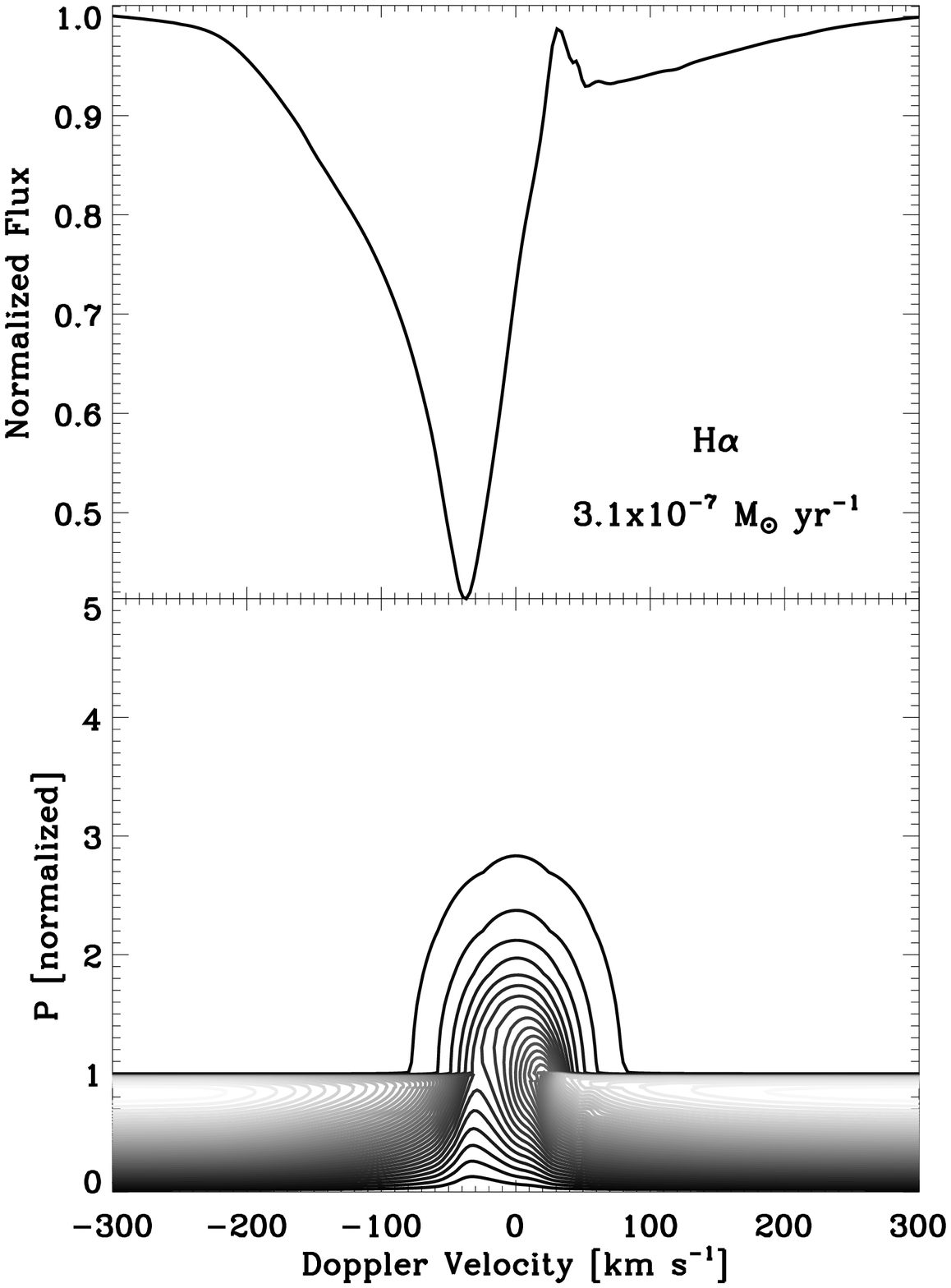}
\includegraphics[width=8.cm]{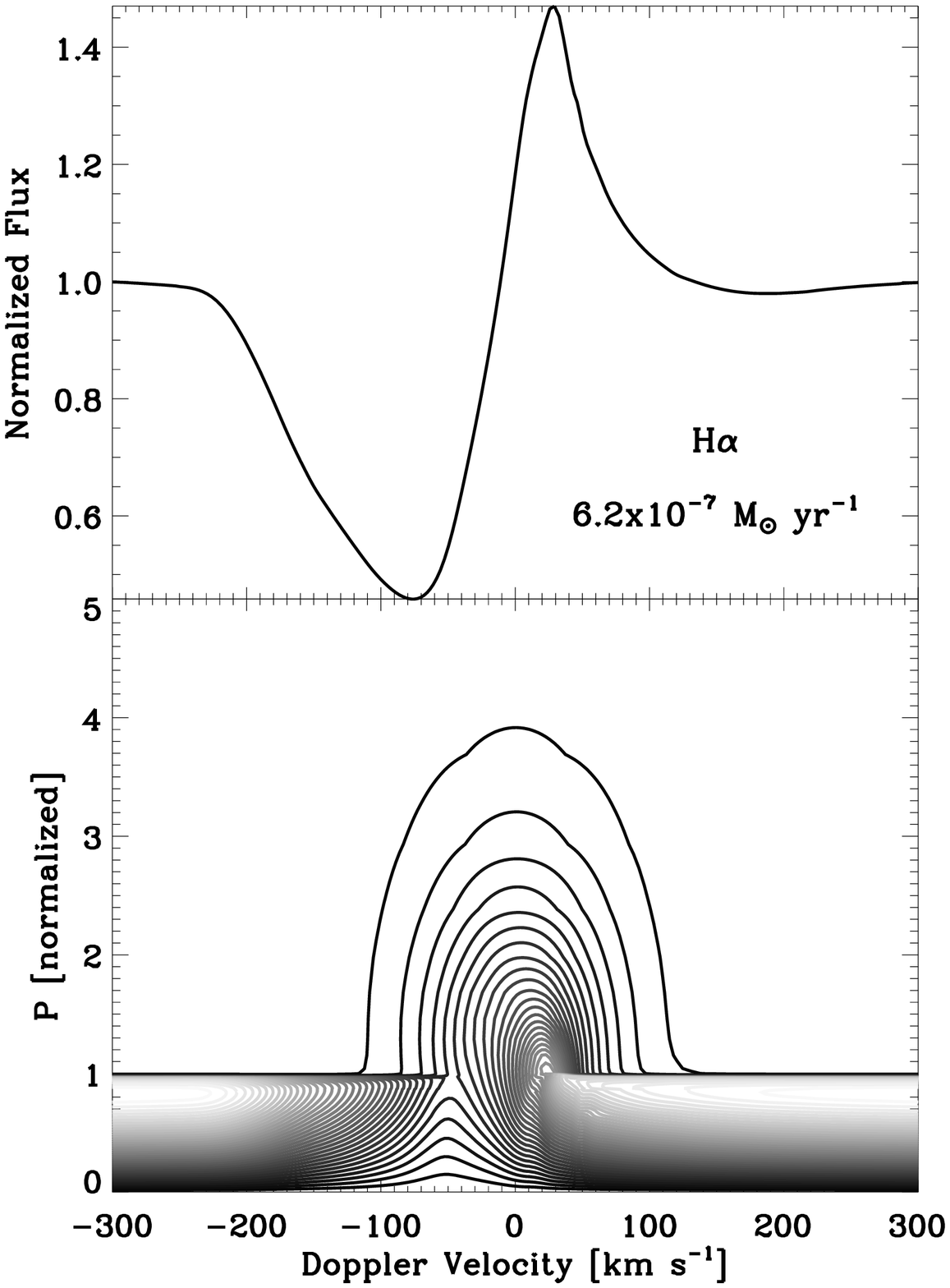}
 \caption[]{
 {\it Left:} In the bottom panel, we show a contour plot of the quantity $P \times I(P)$ as a function of Doppler velocity 
 and impact parameter $P$ for H$\alpha$, computed with {\sc CMFGEN} using the stellar parameters of Deneb 
 and a mass-loss rate value of  3.1$\times$10$^{-7}$\,\msunyr,  corresponding to the best-fit model. This figure illustrates the distribution of intensity with impact parameter, and in particular serves to infer the regions that contribute significant flux to the line, and hence the spatial extent of the H$\alpha$ line-formation region. In the top panel, we show the corresponding normalized synthetic flux in H$\alpha$.
{\it Right:} Same as left, but now for a mass-loss rate value of 6.2$\times$10$^{-7}$\,\msunyr. Notice
the sizable change in profile morphology, echoing the change in the extent of the line-formation region.
\label{fig:masslossDeneb}
}
\end{figure*}

\section{Comparison with radiative transfer models}
\label{sec:model}

\begin{figure*}
 \centering
\includegraphics[width=8cm,angle=0]{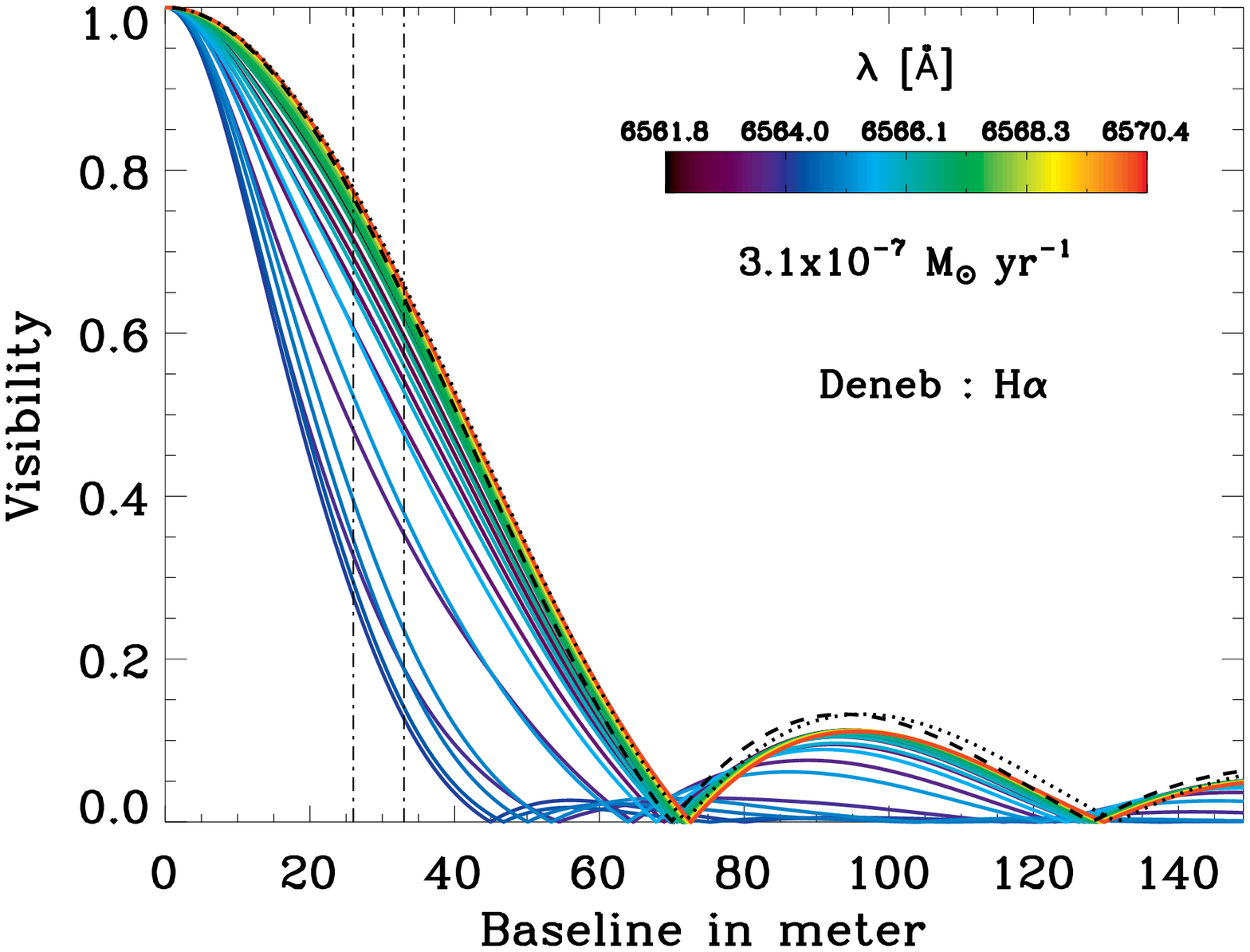}
\includegraphics[width=8cm,angle=0]{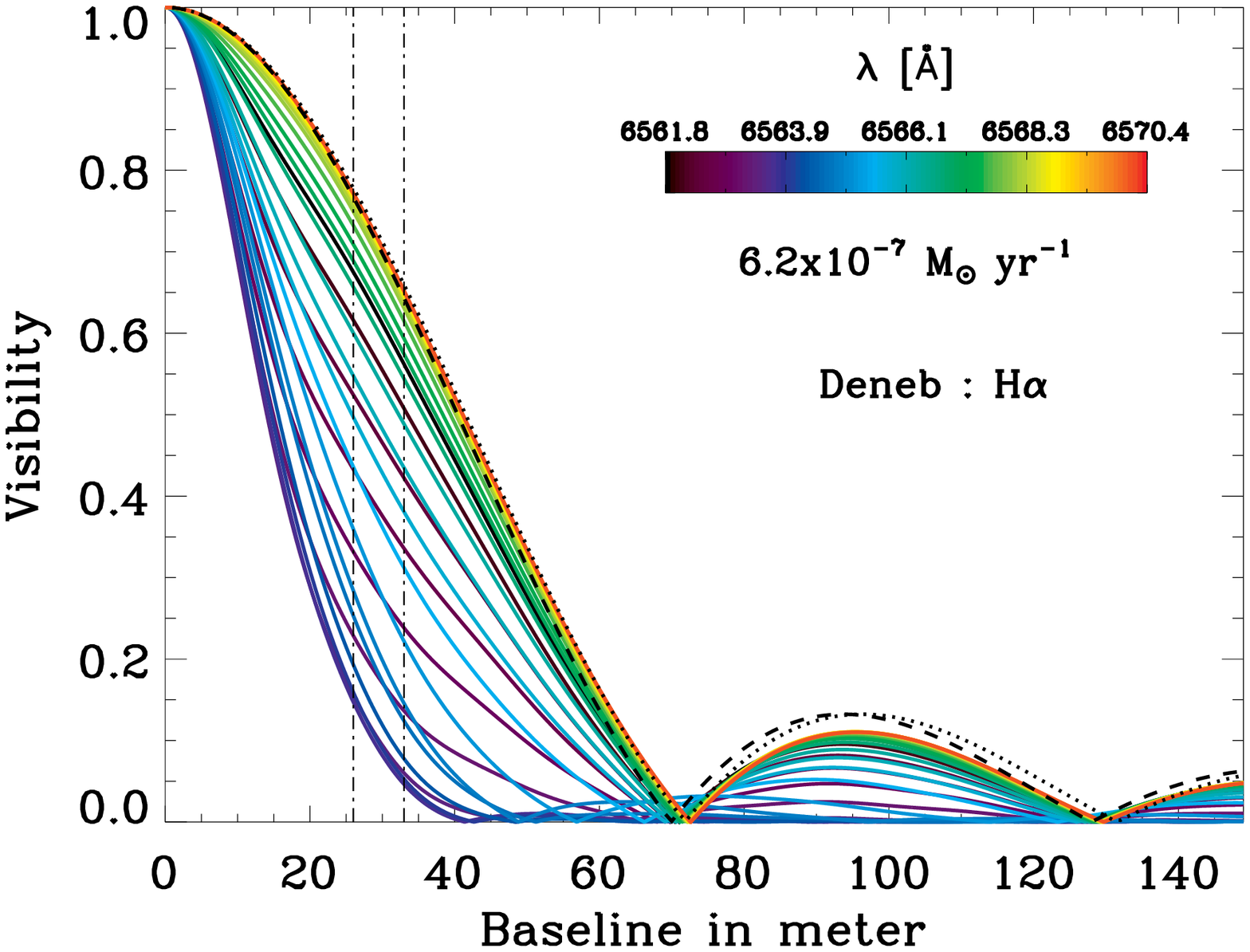}
\includegraphics[width=8cm,angle=0]{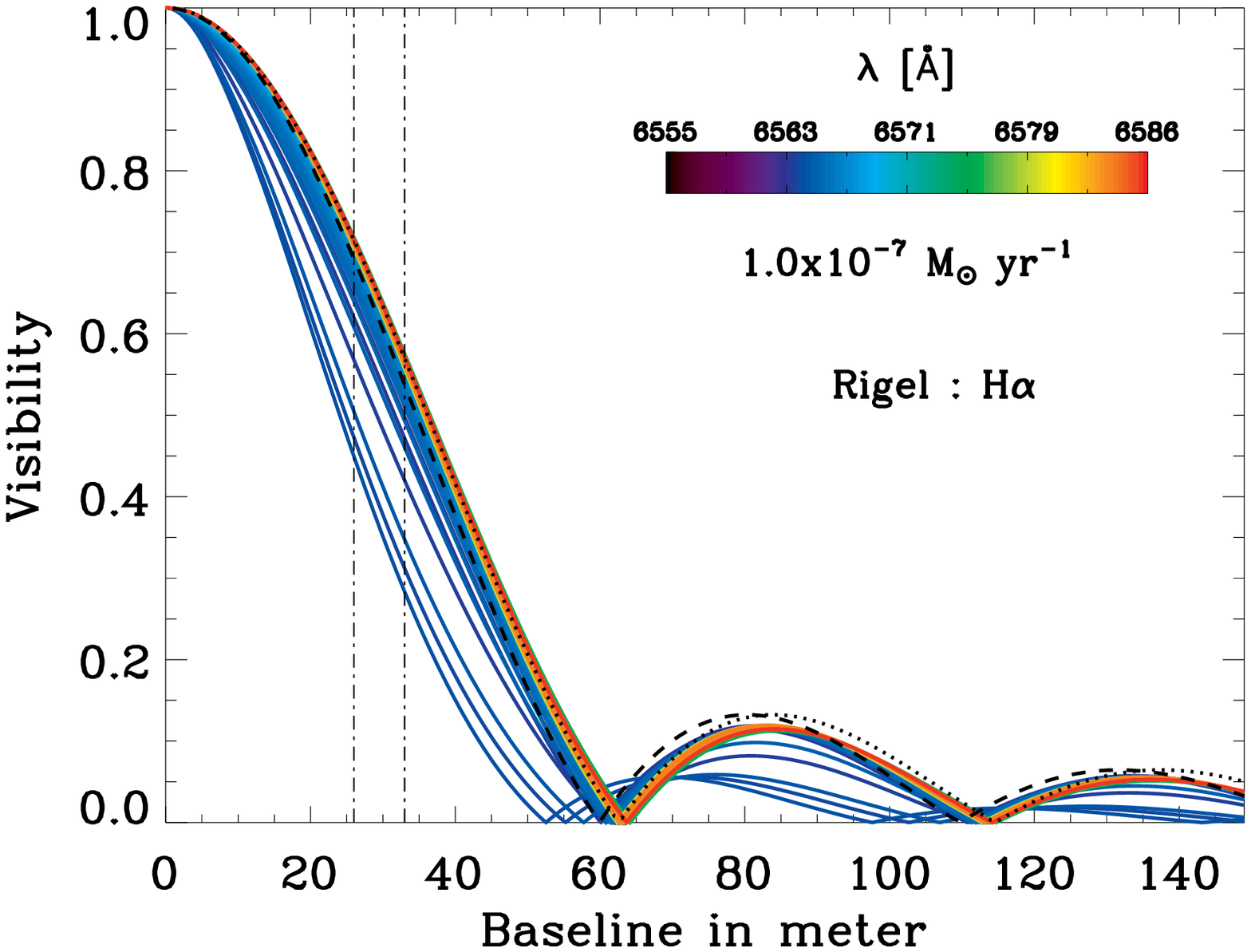}
\includegraphics[width=8cm,angle=0]{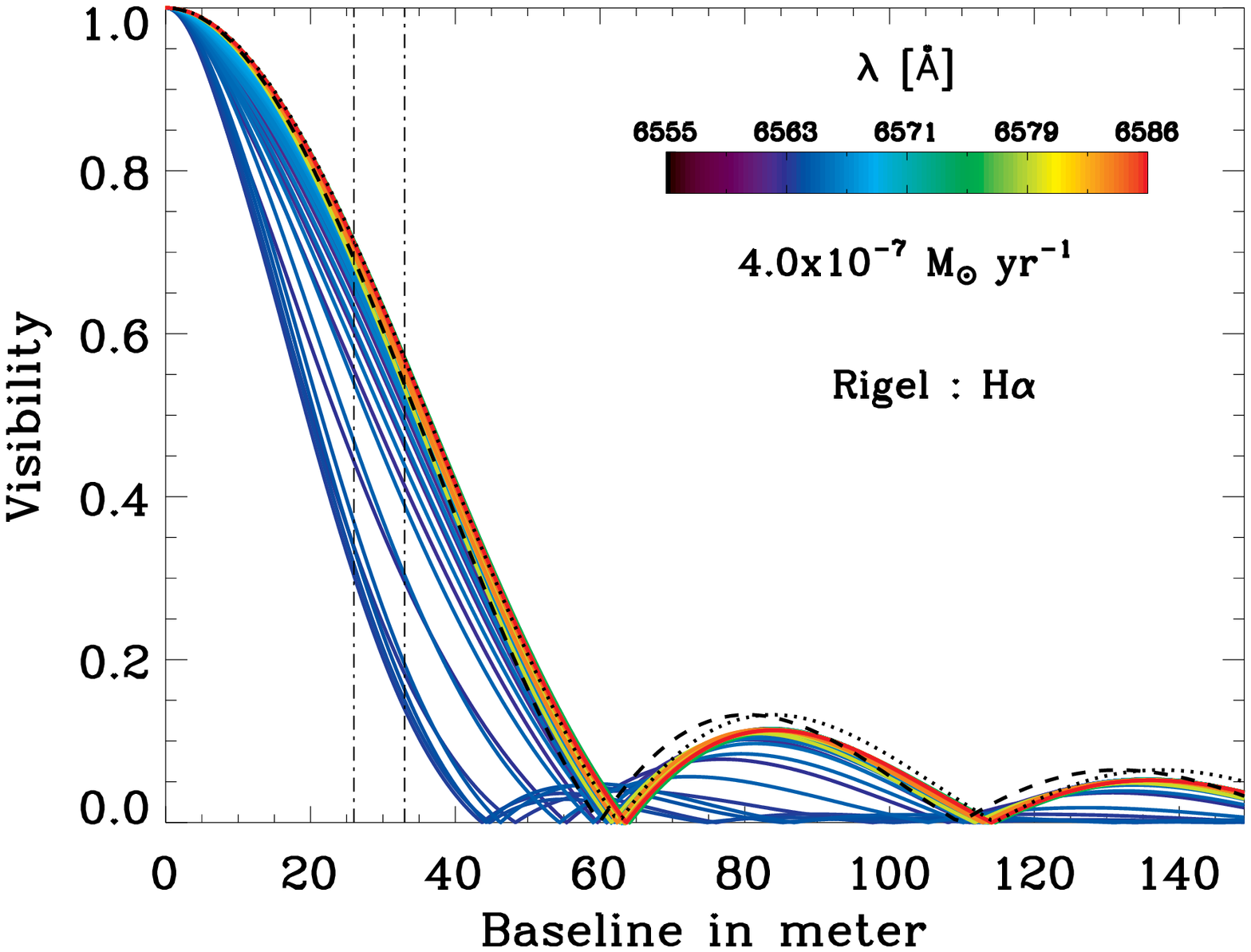}
 \caption[]{{\bf Top:} Theoretical visibility curves for Deneb, computed at selected wavelengths 
 in the range $\sim$6562--6571\AA, and thus containing the H$\alpha$ line, and for 
 mass-loss rate values of 3.1$\times$10$^{-7}$\,\msunyr\ (left) and 
 6.2$\times$10$^{-7}$\msunyr\ (right). 
 {\bf Bottom:} Same as top, but now for Rigel and using mass-loss rate values of 1 (left) and 
 2$\times$10$^{-7}$\,\msunyr\ (right).  The dotted and dashed lines correspond to the best UD curves fitting 
 the visible and the near-IR continua, respectively. The range of baselines of the VEGA/CHARA is indicated by 
 two vertical dash-dotted lines.
\label{fig:CMFGEN}} 
\end{figure*} 

The BA supergiants, especially nearby ones, have been extensively studied with sophisticated 
radiative-transfer codes. It is not in the scope of this paper to determine more reliably the fundamental parameters of 
Deneb or Rigel. However, we wish to investigate several important issues related to the putative effects of the wind on the 
interferometric observables. Deneb has been observed by several northern-hemisphere interferometers in both the visible 
and the near-IR. 
Rigel is currently monitored with VEGA/CHARA in the northern hemisphere, as well as in the southern hemisphere 
with the VLTI.
Hence, in this study, we used the stellar parameters obtained by Przybilla et al. (2006) and Schiller \& Przybilla (2008)
for  Rigel and Deneb, respectively. After developing a convergent model for a reference mass-loss rate value at which
H$\alpha$ is predicted in absorption, we then increased the mass-loss rate (keeping the other stellar parameters fixed)
until the H$\alpha$ line exhibited a well-developed P-Cygni profile.
In this way, we explored the followings questions:
\begin{itemize}
\item To what extent the angular diameter inferred from optical and near-IR continua is sensitive to changes
in mass-loss rates?
\item Are the H$\alpha$, Pa$\beta$, and Br$\gamma$ line-formation regions reproduced by radiative-transfer
simulations accounting for a wind? These lines form in different parts of the wind and can thus be used
simultaneously to constrain its properties.
\end{itemize}

In this paper, we focus on the H$\alpha$ line because it can be observed with VEGA/CHARA. A similar study 
for Pa$\beta$ and Br$\gamma$ line-formation regions is postponed to another paper in preparation.

\subsection{Numerical simulations}

Our radiative-transfer calculations were carried out with the line-blanketed non-LTE 
model-atmosphere code {\sc CMFGEN} \citep{1998ApJ...496..407H, 2005A&A...437..667D},
which solves the radiation-transfer equation for expanding media in the comoving frame, 
assuming spherical symmetry and steady-state, and under the constraints set by the radiative-equilibrium 
and statistical-equilibrium equations. It treats line and continuum processes, and regions of both small and high velocities (relative to the thermal
velocity of ions and electrons). Hence, it can solve the radiative-transfer problem for both O stars, 
in which the formation regions of the 
lines and continuum extend from the hydrostatic layers out to the
supersonic regions of the wind, and Wolf-Rayet stars, in which lines and
continuum both originate in regions of the wind that may have reached
half its asymptotic velocity.

\begin{figure*}
 \centering
\includegraphics[width=8.5cm]{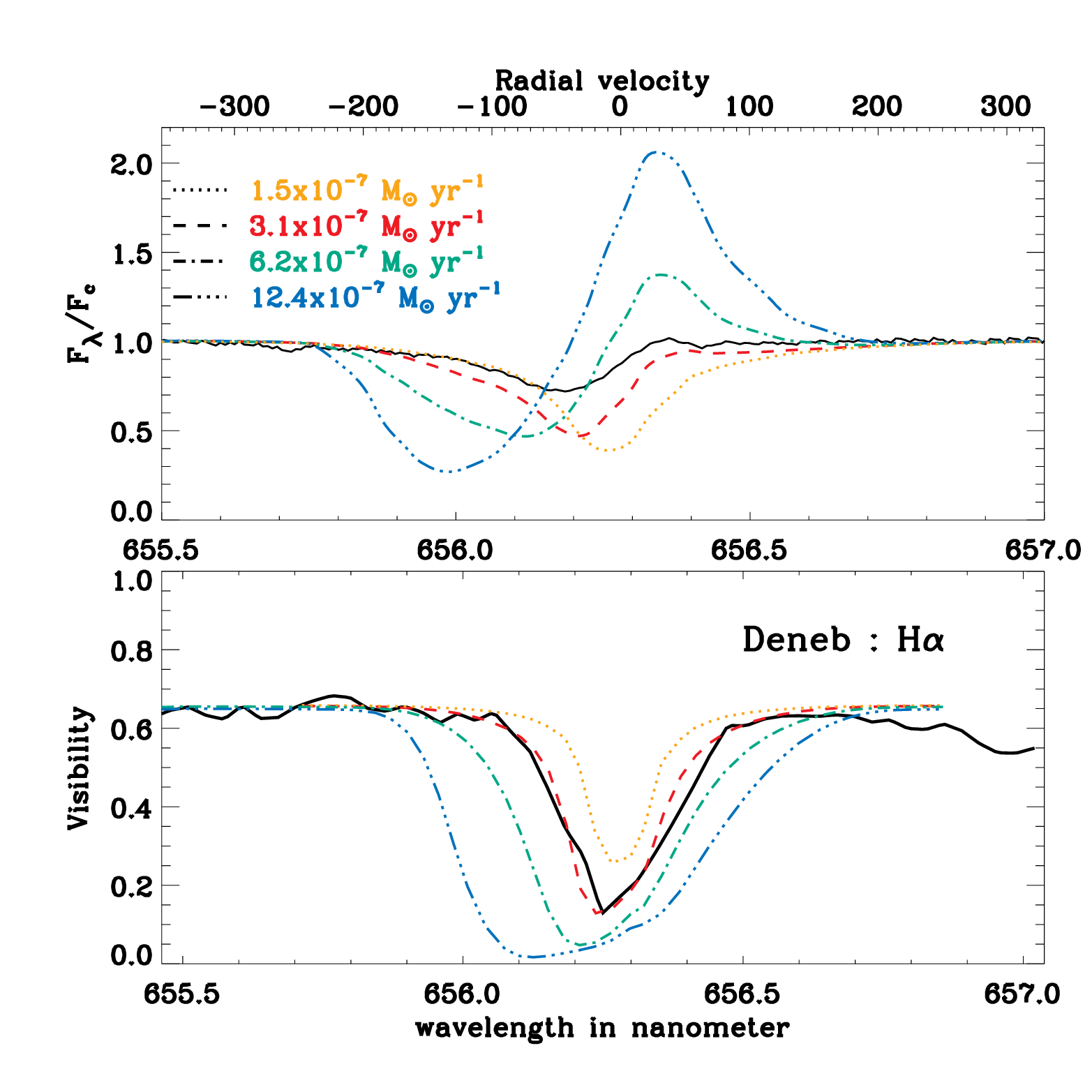}
\includegraphics[width=8.5cm]{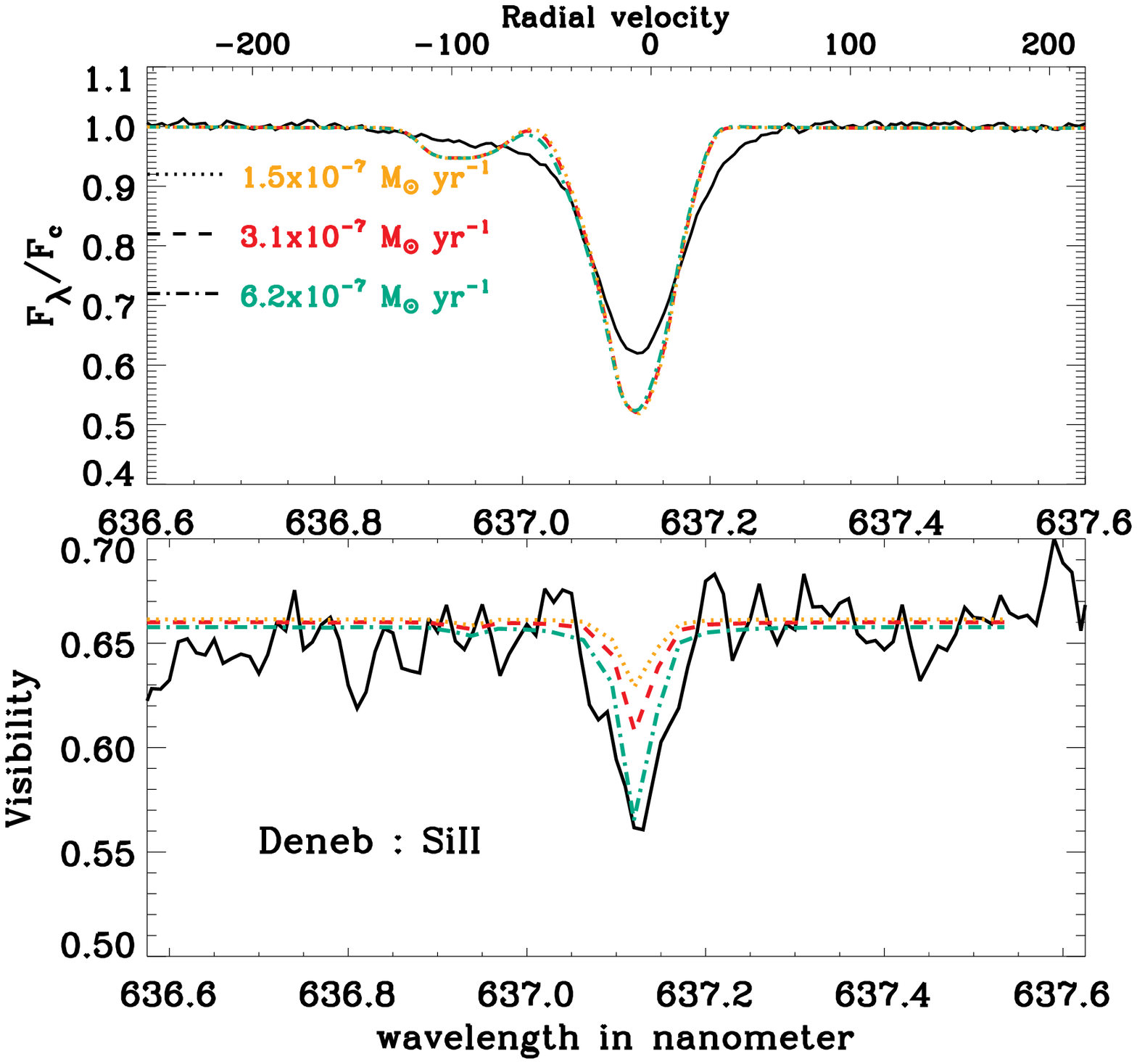}
 \caption[]{
 {\it Left:} In the upper panel, we compare the H$\alpha$ line profile obtained in July 2008 (black; rectified) of Deneb and the 
corresponding model predictions for wind mass loss rate of 1.5 (orange), 3.1 (red), 6.2 (green) and 
12.4$\times$10$^{-7}$\,\msunyr\  (blue). In the lower panel, we give the corresponding dispersed visibility for
a nominal baseline of 33\,m.
{\it Right:} Same as before, but now for Si{\sc ii}\,6371\AA.
\label{fig:cmfgenComp}}
\end{figure*}

We used the stellar parameters inferred by Schiller \& Przybilla (2008) for Deneb, and 
by Przybilla et al. (2006) and  \citet{2008A&A...487..211M} for Rigel.
The Mg{\sc ii} resonance lines suggest terminal wind speeds of  $\sim$240\,\kms 
for Deneb (Schiller \& Przybilla 2008) and $\sim$230\,\kms for Rigel
(Kaufer et al. 1996b). Their projected rotational velocities 
are low, i.e. 20$\pm$2\,\kms and 36$\pm$9\,\kms\, respectively.

Even for sources as ``close" as Deneb and Rigel, the distance estimates remain quite inaccurate. Schiller \& Przybilla (2008) 
derived a luminosity of 1.96$\times$10$^5$\,\lsun\ for Deneb, using a distance of 802$\pm$66\,pc (assuming that 
Deneb is a member of the Cyg OB7 association), while the one derived from Hipparcos \citep{2007A&A...474..653V} 
is significantly  smaller (i.e., 432 $\pm$61\,pc), which infers a luminosity estimate of 5.5$\pm$1$\times$10$^4$\,\lsun. 
This reassessment directly scales down by almost a factor two the linear scales of the size parameters. Furthermore, 
recalling that the mass-loss rate scales with the luminosity, this should imply a weaker steady-state wind mass loss for Deneb. 
The corresponding uncertainties are considerably lower for Rigel, which is much closer than Deneb. 

By adopting values these parameters from previous works, we find that the synthetic spectra computed
by {\sc CMFGEN} agree favorably with the observations. We thus adopt these stellar parameters and vary the mass-loss rate value
to assess the impact on the spectrum and in particular H$\alpha$ and Si{\sc ii}\,6347--6371.4\AA.
In practice, we explore the effect of a wind mass-loss rate of 1.55, 3.1, 6.2, and 12.4$\times$10$^{-7}$\,\msunyr\  for Deneb,
and values of 1, 2, 4, and 8$\times$10$^{-8}$\,\msunyr\ for Rigel.
A typical effect that appears in theoretical models is illustrated in the bottom panels of Fig.~\ref{fig:masslossDeneb}, 
where we show the distribution of the emergent intensity $I$ (scaled by the
impact parameter $p$), as a function of Doppler velocity and $p$. This type of illustration was introduced 
by  \citet{2005A&A...437..667D} to study line formation in hydrogen-rich core-collapse supernova ejecta. 
Here, it provides a measure of the extent of the line-formation region of H$\alpha$ relative to the neighboring
continuum and the sites where most of the emergent photons originate. Comparing the left and right
panels suggests that a variation by a factor of two in the mass-loss rate value leads to sizable changes 
in the extent of the line-formation region, and correspondingly, large changes in the observed line profile.
While spectroscopy is sensitive to the latter, interferometry is sensitive to the former. Below, we describe
the interferometric signals associated with these intensity maps computed with {\sc CMFGEN} and 
produced using the stellar parameters suitable for Rigel or Deneb, and various mass-loss rates tuned
to match observations. This is illustrated in Fig.\ref{fig:CMFGEN} in the case of Deneb and Rigel, using two different mass-loss rates. The visibility curves for various spectral channels in the vicinity of H$\alpha$ are plotted as a function of the spatial frequency. The differential visibilities observed by VEGA are generating by plotting the different value for a given projected baseline (between 27m and 33m).

\subsection{Visible and near-IR continuum forming regions}
\label{contVar}

In Fig.\ref{fig:UDcontinuum}, we show a zoom of the 0.6\,$\mu$m and 2.2\,$\mu$m squared-visibility curves of Deneb 
in the second lobe with the various mass-loss rate values used in this paper. The visible continuum appears to be far 
more sensitive to such changes than the near-IR continuum. We note that these visibility curves are computed 
for a spectral resolution of 30 000, and cannot be directly compared with the broadband measurements 
of FLUOR/CHARA over the full $K'$-band \citep{2008poii.conf...71A}.

Doubling the mass-loss rate from 3.1 to 6.2$\times$10$^{-7}$\,\msunyr\ does not significantly alter the 
optical thickness of the wind. The angular diameter of the star in the continuum near H$\alpha$ 
determined by a UD fitting of the visibility curve does not therefore change by more than 2\%. However, we note that the second 
lobe is significantly affected. This conclusion is also true in the near-IR.
All models also show that the near-IR UD angular diameter is systematically larger than the visible one
by $\sim$2.5-3\%, due to the increasing continuum opacity with wavelength (free-free and bound-free processes). 
The near-IR models were scaled to match the accurate 
CHARA/FLUOR value of 2.363\,mas, and this implies that the expected UD diameter in the 
H$\alpha$ overlapping continuum is $\sim$2.31$\pm$0.02\,mas. This expectation agrees with the NPOI measurements of 
2.34$\pm$0.05\,mas at 0.8$\mu$m and 2.26$\pm$0.06\,mas at 0.55\,$\mu$m.

For Rigel, the difference between the near-IR UD angular diameter and its visible counterpart is increased slightly to reach 
$\sim$3.0-3.5\%. Using the FLUOR/CHARA measurement of 2.758\,mas as reference, this would imply a diameter 
of about 2.64\,mas in the visible. In the near-IR, some instruments such as FLUOR/CHARA 
have an accuracy often better than 1\% depending on the atmospheric conditions, and the ability to observe 
with long baselines. The interpretation of the observations obtained with this instrument might be affected 
by a mass-loss rate variation in the form of localized inhomogeneities, but probably on a smaller scale than an instrument with a similar accuracy operating in the visible..

 The second lobe of the visibility curve is far more sensitive to any fluctuation of the mass-loss rate in the visible than the infrared. The reasons are twofold: a higher sensitivity to limb-darkening effects in the visible, and a more extended continuum-formation region, despite the very limited amount of flux involved (the wind remains in any case optically thin). This can be seen in Fig.\ref{fig:UDcontinuum} as the second lobe of the visibility at 2.2$\mu$m is closer to the uniform disk model than the one at 0.6$\mu$m. Balmer bound-free cross-sections increase from 400 to 800nm. This causes the continuum photosphere to shift weakly in radius across this wavelength range and also alters the limb-darkening properties of the star. This effect might also be discernible in the MarkIII data \citep[ Sect.\ref{sec:diam}]{2003AJ....126.2502M}, Deneb appearing smaller at wavelengths close to the Balmer jump, although the baselines were too short to probe the second lobe of the visibility curve. An interferometer able to resolve a hot star up to the second lobe in the visible is very sensitive to the mass-loss rate, even in the case of a very weak wind. In the near-IR, the free-free emission strengthens. Free-free opacities increase at longer wavelengths, increasing the photospheric radius relative to that measured in the visible. Changing the mass-loss rate does not significantly affect the limb-darkening and the temperature scale near the star, thus the location of the extended emission. As a consequence, the free-free emission causes a larger diameter in the near-IR, independently to first order of the mass-loss rate.

\subsection{Dependence of the H$\alpha$ and Si{\sc ii} lines on the mass-loss rate}

We then computed the spectrum and the dispersed visibilities in the H$\alpha$ line 
at a spectral resolution R=30 000. We were unable to perform a satisfactory fit to the H$\alpha$ profile for either Deneb or Rigel, 
the absorption component being systematically too deep (Fig.\ref{fig:cmfgenComp}). This conclusion was reached in many studies, 
and these profiles could not be reproduced in terms of spherically-symmetric smooth wind models (see for instance 
Fig.~7 in \citet{2008A&A...487..211M}, \citet{2002ApJ...570..344A}, and \citet{2008A&A...479..849S}). In this temperature 
regime, the models systematically predict profiles in absorption partly filled-in by wind emission, hence  
only lower limits to the mass-loss rate can be derived by fitting the H$\alpha$ line.

However, this discrepancy is mitigated by the quality of the visibility fit shown in the upper panel of Fig.\ref{fig:cmfgenComp}, 
which is by far less sensitive to small absorption effects along the line-of-sight. Changing the mass-loss rate by a large factor of 2--4
has a dramatic impact on the H$\alpha$ line-formation region, changing the spectrum appearance and the dispersed visibility curve. 
When using a baseline in the range 26-33\,m, there is a relationship between the mass-loss rate and the FWHM of the visibility 
drop in the line that can be approximated by a second-order polynomial in the range considered 
(1.55 to 12.4$\times$10$^{-7}$\msunyr). One can see in Fig.~\ref{fig:massloss} that the nominal model of Deneb with a mass-loss 
rate of 3.1$\times$10$^{-7}$\,\msunyr\ fits the observed visibility curves well. The model FWHM of the visibility is 89\,\kms\ and 
the observed FWHM is 98$\pm$3\,\kms. Given the FWHM-mass-loss relationship, this would correspond to a mass-loss rate 
in the range 3.7$\pm$0.2$\times$10$^{-7}$\,\msunyr. For Rigel, a similar relationship provides a mass-loss rate 
in the range 1.5$\pm$0.4$\times$10$^{-7}$\,\msunyr.
We have studied the variations in 2008 and 2009 of the visibility profile without detecting any significant 
variation in the visibility profile, which, translated in a mass-loss rate variation, suggests that the mass-loss rate 
has not changed by more than 5\%. 

By comparison, the Si{\sc ii}\,6371\AA\ line is less sensitive to any variation in the mass-loss rate (Fig.\ref{fig:cmfgenComp}). 
The FWHM of the simulated line weakly increases from 40 to 43\,\kms, compared to the 2008 VEGA/CHARA measurement of 
55$\pm$3\,\kms. The FWHM of the corresponding model visibility dip are independent of the mass-loss rate, being unchanged at the value 
19.5\,\kms and lower than the measured mean value of 29$\pm$5\,\kms\ for the 2008 observations. The depth of the visibility dip  
fits more accurately the curve with a mass-loss rate of 6.2$\times$10$^{-7}$ rather than the nominal value of 3.1$\times$10$^{-7}$\,\msunyr\ inferred from the 
H$\alpha$ line. It is not possible to establish whether this issue is related to a particular event occurring during the 2008 observations, 
or a permanent situation, or even a bias in the radiative-transfer model.

\begin{figure}
 \centering
\includegraphics[width=8.7cm]{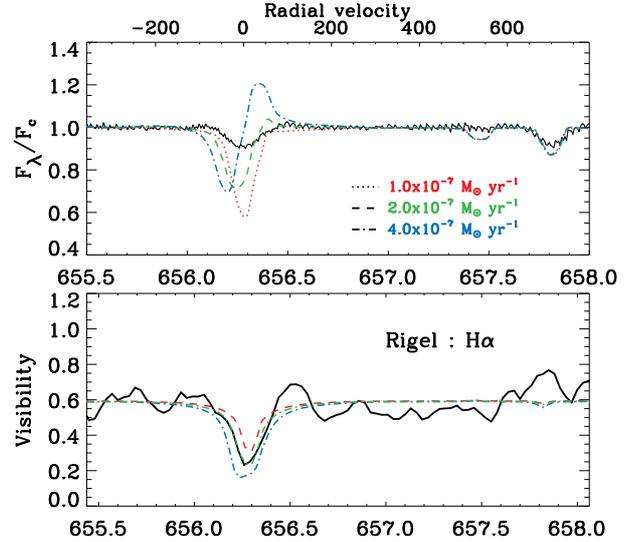}
 \caption[]{Same as the upper panel of Fig.~\ref{fig:cmfgenComp}, but now for Rigel. 
\label{fig:cmfgenCompRigel}}
\end{figure}

Despite the lower quality of the interferometric data on Rigel, one can see in Fig.\ref{fig:cmfgenCompRigel} that the model provides 
a reasonable fit. We note that the C{\sc ii} lines at 6578\AA\ and 6583\AA\ in the spectrum of Rigel are also slightly affected by the wind. 
Their visibility for the VEGA baselines are about 0.5-1\% lower than the nearby continuum (using our model predictions), 
depending on the mass-loss rate. 
Detecting such a weak signal would require that the accuracy on the differential visibility is about 2-3 times better than the current instrument performance.

\begin{figure}
 \centering
\includegraphics[width=8.cm]{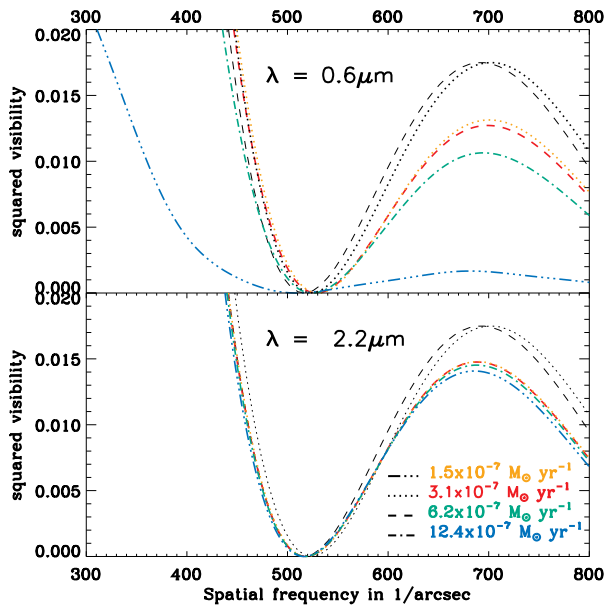}
 \caption[]{ Synthetic continuum squared-visibility curves at 0.6$\mu$m (top) and 2.2$\mu$m (bottom)
 as a function of spatial frequency, and given for wind mass-loss rates of 1.5 (orange), 3.1 (red),
 6.2 (green), and 12.4$\times$10$^{-7}$\,\msunyr\  (blue). The dotted black line corresponds to a uniform disk of 3.32mas, and the dashed black line to a uniform disk of 3.36mas. 
\label{fig:UDcontinuum}}
\end{figure}

\begin{figure}
 \centering
\includegraphics[width=8.cm]{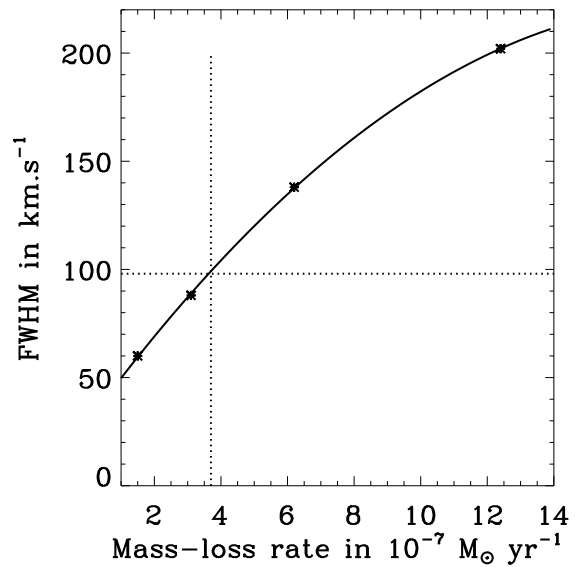}
 \caption[]{Variation in the FWHM measured from the H$\alpha$ visibility curves computed with the
 intensity maps produced by {\sc CMGEN}, shown here as a function of mass-loss rate and adopting
 the stellar parameters suitable for Deneb. The adopted baseline is 33\,m.  
  For a high mass-loss rate value, the H$\alpha$ line-formation region is fully resolved, and the FWHM 
 of the visibility curve is close to the wind terminal velocity. The dotted lines indicate the mass-loss rate inferred from the VEGA/CHARA measurements.\label{fig:massloss}}
\end{figure}

\subsection{A rough investigation of the morphology of the line-formation regions}

The variability in the differential phase signal observed is a clear sign of the stochastic activity of the wind of Deneb. Even though the data secured are partial, one can note that no differential phase signal exceeds 30\deg, which is impressively large considering that the baseline is limited. In 2008 and in October 2009, the  differential phase signal was able to be observed throughout the full line, at a level above 30\deg. This signal did not appear to be due to an increased level of stochastic 'noise' expected from the signature of small, localized clumps, but exhibited a well-structured signature. At other times, the differential signal was only observed close to $\pm v sini$, and some patterns were reminiscent of the S-shape signature, originating in a rotating structure. 

\begin{figure*}
 \centering
\includegraphics[width=18.5cm]{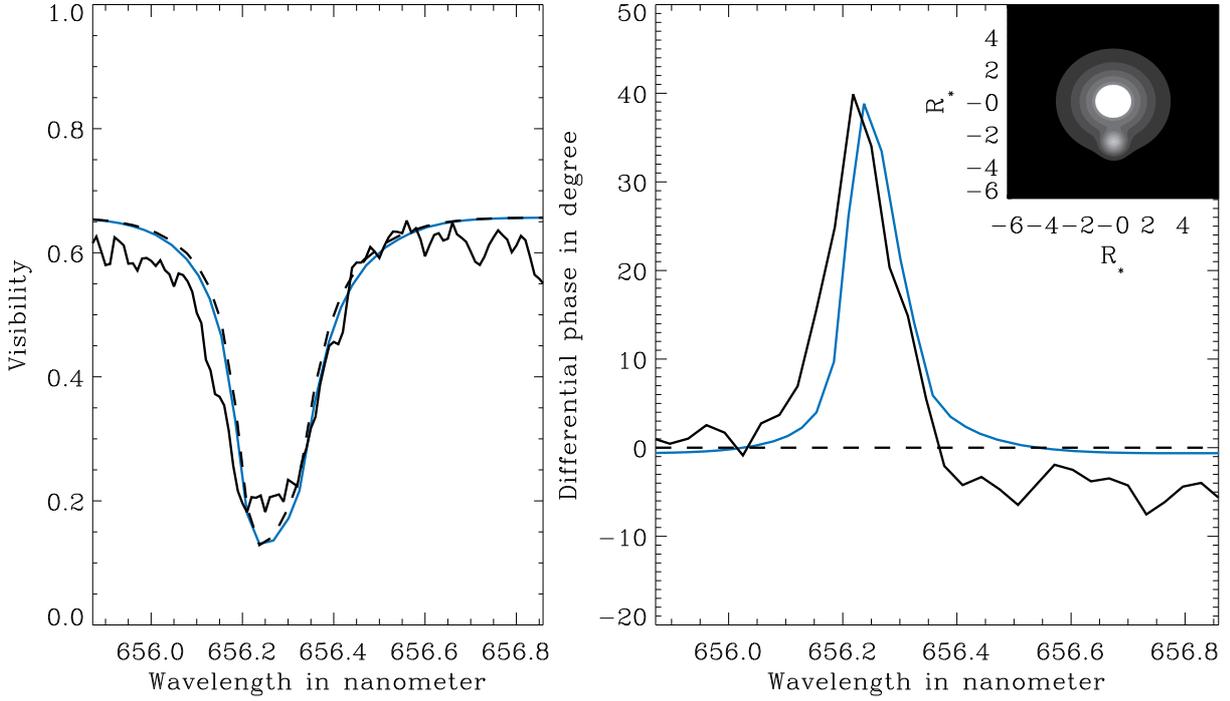}
 \caption[]{Comparison on the unperturbed model (dashed lines) with the perturbed model (blue lines, as explained in the text). The thick lines are the observation of July 2008. The parameters of the perturbation are the followings: flux contrast of 20, FWHM of 0.5 stellar radius, and position at 2.8 stellar radius south of the star. The insert shows the square root of the perturbed model at the wavelength corresponding to the core of the line, where the signal is maximum.  
\label{fig:cmfgenPerturb}}
\end{figure*}

The radiative transfer models demonstrate that at the inferred mass-loss rate of Deneb, significant emission originates in a circumstellar region of up to 2-4 stellar radii, and that the level of this extended emission depends strongly on the mass-loss rate. We tried to investigate the differential phases with ad hoc approaches using the models as a basis to generate the perturbation. Differential visibilities and phases provide contradictory and stringent constraints. On the one hand, a strong differential phase signal is observed, but on the other hand, the differential visibilities did not vary by more than 5\% over the two years of sparse observations. This restricted considerably the size, flux, and location of the perturbation. 
The emergent flux of the best-fit model was perturbed using a 2D Gaussian as a weighting function. The parameters of the perturbation were its location, FWHM and flux ratio compared to the unperturbed model.  The best parameter range was reached for a perturbation located at 2-3 stellar radii from the star, with a FWHM of 0.3-0.7 stellar radii, and a flux contrast of 10-25. One can see in Fig.\ref{fig:cmfgenPerturb} that the differential visibilities of the perturbed and unperturbed models are similar, whereas a significant differential phase signal is observed. The baseline is aligned in the direction of the perturbation observed in July 2008, i.e. north of the star. This model is also compatible with the differential phases observed with the two other baselines (Fig.\ref{fig:vis}). Placing the perturbation farther away leads to a phase signal that is far weaker than to the observations, since the line-forming region of the unperturbed model naturally ends at 5-6 stellar radii. This is a limitation of this ad hoc approach. Making the perturbation larger in size, even with a much lower flux contrast leads to a noticeable change in the differential visibilities. The same consequence is reached when putting the perturbation too close to the star. Finally, if the perturbation is too small (i.e. FWHM $\leq$ 0.1 mas), the flux contrast required to reach the observed phases is unrealistically high, and such that a significant 'binary'-like perturbation in the visibilities should be observed. Finally, this perturbation study suggests that a large temporal variability in the visibility and phase is expected in the visible at high spatial frequencies corresponding to the second lobe of the visibility curve.

\citet{1976MNRAS.174..335C} and \citet{1979MNRAS.186..245M} did  
not detected any significant variation in the polarization of the H$\beta$ and H$\alpha$ lines in Deneb and Rigel, at a typical level of about 0.1\%. \citet{1986ApJ...302..403H} performed an intensive broadband monitoring in $B$-band polarization of Rigel, revealing a variability at a typical level of 0.2\%. 
The variability pattern in the $Q-U$ plane suggested that the ejection of material was not limited to a plane, and non-radial pulsation were thus proposed 
as the root cause of these localized ejections. This might be caused by the limited amount of intensive observations. This might also be a consequence, in the case of Deneb, of a small inclination of the rotation axis on the sky. Any large-scale structures orbiting in the equatorial plane would generate a significant polarization variability, without any preferential direction. In addition, no direction appears to be preferred by our observation, though we emphase that such a conclusion is very limited given the stringent limitations in the $uv$ coverage.

\section{Deneb as a fast rotator}

\citet{2008poii.conf...71A, 2006AAS...208.0601A} presented evidence that Deneb is a fast rotator, based on 24 high accuracy 
FLUOR/CHARA visibilities in the $K'$-band with projected baselines ranging from 106\,m to 310\,m, sampling the first and second lobes 
of the visibility curve. They detected slight departures from a purely spherical model at a level of about 2\% in the near-IR, a discovery 
that had not been anticipated for an AB supergiant. They tentatively fitted the data with a rotating model atmosphere, and demonstrated that despite 
the low $v\sin i$ of the star, a model at half critical speed, seen nearly pole-on may account for the 
interferometric observations.
A by-product of this fitting process is the determination of crucial parameters in this context, namely
the inclination, estimated to be $i$=30\deg, and the orientation on the sky of the rotation axis found to be at about P.A.=150\deg\
 east from north. These findings have potentially important implications. In a general context, Deneb would be the first AB supergiant 
 proven to be the descendant of a fast rotator. Support for this interpretation was provided by \citet{2008A&A...479..849S}. To more clearly  
 interpret the pronounced mixing signature with nuclear-processed matter, these authors proposed that Deneb was 
probably a fast rotator initially, and is currently evolving to the red-supergiant stage.

In the context of our observations, the consequences on the H$\alpha$ line-formation region must be evaluated. The H$\alpha$ line 
may be affected by a moderate change in wind properties, such as those that occur due to a latitudinal 
variation in the effective temperature of the star, estimated to be $\sim$700\,K \citep{2006AAS...208.0601A}. 
Moreover, as stated by \citet{2008A&A...479..849S}, hydrogen lines are mainly sensitive to variations in $\log g$.

Neither the VEGA/CHARA data nor the theoretical study presented in this paper can provide definitive support for, or exclude, this interpretation. 
The H$\alpha$ variability observed by VEGA/CHARA is related to localized inhomogeneities in the wind of this star. Is not impossible that
that these inhomogeneities may affect the continuum forming region in the K band, although is has been shown in Sect.~\ref{contVar} that 
large variations in mass-loss rate are required to significantly affect the second lobe of the visibility curves. Yet, this conclusion is based on the ideal case of a spherical model, although given that the wind is very optically thin in the continuum, those inhomogeneities should not significantly alter the properties
of the continuum-formation regions.

The 2008 data arguably provide some support for the fast rotator model. We note that the S1S2 projected baselines were roughly 
aligned (range of -45/+15 degrees) within the direction of the asymmetry found with CHARA/FLUOR at 150\deg, i.e. the direction 
of the pole in the fast rotator model \citep{2006AAS...208.0601A}. On the other hand, the H$\alpha$ differential phases are observed 
to increase from a baseline roughly aligned to the pole direction at PA=150\deg\ . 
This may imply that the asymmetry is greater in a direction perpendicular to the pole. On the other hand, the Si{\sc ii} line dispersed visibilities seem deeper in the direction of the pole, suggesting a more extended line-formation region. Moreover, the high mass-loss rate inferred from the fit to the visibilities through the SiII line may also be an indication of a co-latitude dependence of the mass-loss rate. That no 
evidence of rotation was detected in the differential phases is an additionnal argument for a low $v\sin i$, and therefore a nearly 
pole-on configuration for Deneb. These limited observations cannot provide definite conclusions, and the fast-rotator 
interpretation still needs to be investigated, both theoretically and observationally. This could be done, for instance, by repeating the 
FLUOR/CHARA observations to check whether the near-IR interferometric signal has changed or not, or by devoting a full 
VEGA/CHARA run with more extensive coverage, preferably by using the 3T mode. It would also be of interest to investigate 
the impact of the fast rotation of Deneb on the H$\alpha$ and Si{\sc ii}\,6371\AA\ line-forming regions, using a radiative transfer model of a 
rotating star with a wind.

\section{Conclusion}

We have presented the first high spatial and spectral observations of two nearby supergiant stars Deneb and Rigel. 
The H$\alpha$ line-formation regions were resolved and their angular size was found to be in agreement with an up-to-date 
radiative transfer model of these 
stars. The H$\alpha$ line-forming region appears to be asymmetric and time variable, as expected from the numerous 
intensive spectral monitoring of AB supergiants reported in the literature. However, the time-monitoring of the dispersed visibilities inferred from the H$\alpha$ line 
of Deneb did not provide evidence of mass-loss rate changes above 5\% of the mean rate, and the activity observed can be considered most of the time as a 
second-order perturbation of the wind characteristics. The increased coverage obtained 
in 2008 provides some evidence of a latitudinal dependence of both the H$\alpha$ differential phases and the Si{\sc ii}\,6371 
line differential visibilities. This line was not significantly resolved subsequently. Only two observations of Rigel were secured. 
An S-shape signal was detected in the H$\alpha$ differential phase of Rigel, suggesting that the rotation signal is detected. 
We note that the H$\alpha$ line profile was almost photospheric at the time of our observations. Observations with a large $uv$ 
coverage may provide, in the case of Rigel, the direction of the rotational axis on the sky and its inclination. 

Given the large angular size of Rigel and Deneb, the H$\alpha$ line-forming region is fully resolved by the interferometer 
with baselines longer than 50m, and only the S1S2 baseline of the CHARA array is short enough to permit such an investigation. 
Therefore, it is not easy to significantly increase the observational effort performed on these stars with such stringent observing 
restrictions. An extension of this work is possible for sources with an apparent angular diameter between 0.5 and 1.5\,mas, 
that are large enough, but also bright enough to make use of the highest spectral dispersion of the VEGA instrument (limiting 
magnitude of about 3). This concerns the AB supergiants closer than 1.5 kpc, and the supergiants in the Orion complex 
($d \sim$500\,pc) are in this context of particular interest. Simultaneous 3 telescope recombination would provide much better $uv$ coverage 
than the one presented in this paper.  Another interesting possibility is to add to this study several diagnostic lines such as the 
Ca{\sc ii} infrared triplet (849.8nm, 854.2nm, 866.2nm). These resonance lines are highly sensitive to non-LTE effects arising 
close to the photosphere and may shed some light on the regions were the material is launched.

\begin{acknowledgements}
VEGA is a collaboration between CHARA and OCA/LAOG/CRAL/LESIA that has been supported by the French programs PNPS and ASHRA, by INSU and by the r\`egion PACA. The project has obviously taken benefit from the strong support of the OCA and CHARA technical teams. The CHARA Array is operated with support from the National Science Foundation and Georgia State University. We warmly thank Christian Hummel for having provided the MarkIII data. The referee, Mike Ireland helped us by his useful comments to improve this paper significantly.
This research has made use of the Jean-Marie Mariotti Center \texttt{SearchCal} service \footnote{Available at http://www.jmmc.fr/searchcal}
co-developed by FIZEAU and LAOG, and of CDS Astronomical Databases SIMBAD and VIZIER \footnote{Available at http://cdsweb.u-strasbg.fr/}. M.B.F. acknowledges Conselho Nacional de Desenvolvimento Cient\'ifico e Tecnol\'ogico (CNPq-Brazil) for the post-doctoral grant.
\end{acknowledgements}


\bibliographystyle{aa}
\bibliography{Bib_super} 

\appendix

\end{document}